\documentclass[12pt]{article}

\usepackage[dvips]{graphicx}
\renewcommand{\baselinestretch}{1.75}
\font\scaps=cmcsc10 scaled\magstep1 

\newcommand \be{\begin{equation}}
\newcommand \ee{\end{equation}}
\newcommand \ba{\begin{eqnarray}}
\newcommand \ea{\end{eqnarray}}
\begin{document}

\def\today{\ifcase\month\or
January\or February\or March\or April\or May\or June\or July\or August\or
September\or October\or November\or December\fi
\space\number\day, \number\year}
%

\hfil PostScript file created: \today{}; \ time \the\time \ minutes
\vskip .15in

\centerline {TOHOKU EARTHQUAKE: A SURPRISE? }

\vskip .15in
\begin{center}
{Yan Y. Kagan and David D. Jackson}
\end{center}
\centerline {Department of Earth and Space Sciences,
University of California,}
\centerline {Los Angeles, California 90095-1567, USA;}
\centerline {Emails: {\tt kagan@moho.ess.ucla.edu,
david.d.jackson@ucla.edu}}
\vskip 0.02 truein

\vspace{0.15in}

\noindent
{\bf Abstract.}
We consider three questions related to the 2011 Tohoku
mega-earthquake:
(1) Why was the event size so grossly under-estimated by
Japan's national hazard map?
(2) How should we evaluate the chances of giant earthquakes in
subduction zones? and
(3) What is the repeat time for magnitude 9 earthquakes off
the Tohoku coast?
The ``maximum earthquake size" is often guessed from the
available history of earthquakes, a method known for its
significant downward bias.
We show that historical magnitudes systematically
under-estimate this maximum size of future events, but the
discrepancy shrinks with time.
There are two quantitative methods for estimating the corner
magnitude in any region: a statistical analysis of the
available earthquake record, and the moment conservation
principle.
However, for individual zones the statistical method is
usually ineffective in estimating the maximum magnitude; only
the lower limit can be evaluated.
The moment conservation technique, which we prefer, matches
the tectonic deformation rate to that predicted by earthquakes
with a truncated or tapered magnitude-frequency distribution.
For subduction zones, the seismic or historical record is
insufficient to constrain either the maximum or corner
magnitude.
However, the moment conservation principle yields consistent
estimates: for all the subduction zones the corner
magnitude is of the order 9.0--9.7.
Moreover, moment conservation indicates that variations in
estimated corner magnitude among subduction zones are not
statistically significant.
Another moment conservation method, applied at a point on a
major fault or plate boundary, also suggests that magnitude 9
events are required to explain observed displacement
rates at least for the Tohoku area.
The global rate of magnitude 9 earthquakes in subduction
zones, predicted from statistical analysis of seismicity as
well as from moment conservation is about five per century --
five actually happened.

\vskip .15in
\noindent
{\bf Short running title}:
{\sc
Tohoku earthquake: a surprise?
}

\vskip 0.05in
\noindent
{\bf Key words}:
\vskip .05in
Probability distributions;
Seismicity and tectonics;
Statistical seismology;
Dynamics: seismotectonics;
Subduction zones;
Maximum/corner magnitude.

\section{Introduction}
\label{intro}

The 11 March 2011 Tohoku, Japan magnitude 9.1 earthquake and
the ensuing tsunami near the east coast of Honshu caused
nearly 20,000 deaths and more than 300 billion
dollars in damage, ranking as one of the worst natural
disasters ever recorded (Hayes {\it et al.}, 2011; Simons {\it
et al.}, 2011; Geller, 2011; Stein {\it et al.}, 2011).
The great difference between the expected and observed
earthquake magnitudes contributed to this enormous damage.
The estimated maximum magnitude for the Tohoku area (around
7.7) was proposed in the official hazard map (Headquarters for
Earthquake Research Promotion, 2005; Seismic Activity in
Japan, 2008; Simons {\it et al.}, 2011).

Several other estimates of the maximum size earthquake in the
Tohoku area have been published before 2011.
Ruff and Kanamori (1980, Fig.~A1) suggested, basing their
analysis on historical data, age of subducting plate and
plate-rate systematics, that the maximum magnitude is
8.2-8.35.
Wesnousky {\it et al.}\ (1984, Table~1, Fig.~8)
applied the
characteristic earthquake hypothesis to estimate the maximum
earthquake size and expected occurrence time for these events
in three zones approximately covering the Tohoku earthquake
rupture area.
Nishenko (1991, Fig.~23 and pp.~234-235) defines the magnitude
of characteristic earthquakes in NE Japan, zones $J4-J6$
as 7.1-7.7.
However he does mention an $m8.1$ earthquake in 1611.
Minoura {\it et al.}\ (2001) suggest that the 869 Jogan
earthquake occurred on a fault $ 240 \times 85$~km and was
$m8.3$.
A similar estimate of the Jogan earthquake magnitude is
proposed by Sugawara {\it et al.}\ (2012).
On the basis of the historical and instrumental catalog
analysis, Grunewald and Stein (2006) suggest the maximum
magnitude $m8.4$ for the Tokyo area.
Koravos {\it et al.}\ (2006) provide an estimate $7< m_{\rm
max} <8$, based on historical earthquakes since 599 AD, again
mentioning a few earthquakes with magnitude slightly larger.
Annaka {\it et al.}\ (2007) suggest $m_{\rm max} \approx 8.5$
based on historical earthquakes since 1611.
Rikitake and Aida (1988, Table~1, Fig.~5) did not expect
the Tohoku area tsunami exceeding 7~m, and defined the maximum
magnitude for zones $III-V$ to be within $7.4-7.9$.
``Evaluation of Major Subduction-zone Earthquake(s)" 2008 PDF
file at ``Seismic Activity in Japan" Japanese Web site
(http://www.jishin.go.jp/main/index-e.html) defines
probabilities of major earthquakes in subduction zones.
For the north-east part of Honshu Island, the maximum
magnitudes are in the range 6.8-8.2.
Nanjo {\it et al.}\ (2011) suggest that one should expect
magnitudes ``up to about 8 or larger for [Japan] offshore
events."
These opinions of the Japanese and international researchers
confirm that the maximum earthquake size in the Tohoku area
was dramatically under-estimated before 2011/03/11.

Several quantitative estimates of the maximum possible
earthquakes in the subduction zones had been published before
the Tohoku event (Kagan, 1997; Kagan and Jackson, 2000; Bird
and Kagan, 2004; McCaffrey, 2007, 2008; Kagan {\it et al.},
2010).
In these publications, the upper magnitude parameter was
determined to be within a wide range from 8.5 to 9.6.

Two quantitative methods have been deployed to estimate the
upper magnitude limit: a statistical determination of
the magnitude--moment/frequency parameters from earthquake
data alone, and a moment conservation principle in which the
total moment rate is estimated from geologic or geodetic
deformation rates (Kagan, 1997).

The distinction between maximum and corner magnitudes refers
to different approaches in modeling the magnitude-frequency
relationship for large earthquakes.
In one approach, the classical Gutenberg-Richter magnitude
distribution is modified by truncating it at an upper limit
the ``maximum magnitude."
In another approach, an exponential taper is applied to the
moment distribution derived from the classical magnitude
distribution.
For the tapered Gutenberg-Richter (TGR) distribution
the ``corner moment" is the value at which the modeled
cumulative rate is reduced to 1/$e$ of the classical rate, and
the ``corner magnitude" is that which corresponds to the
corner moment.
In either approach, the rate of earthquakes of any magnitude
and the total seismic moment rate can be computed from three
observable parameters: the rate of earthquakes at the lower
magnitude threshold, the ``$b$-value" or asymptotic slope of
the magnitude distribution, and the maximum or corner
magnitude.
Conversely, the maximum or corner magnitude may be determined
if the threshold rate, $b$-value, and total moment rate are
known.

We generally shun the use of ``maximum magnitude" because
there is no scientific evidence that it really represents a
maximum possible magnitude, although it is frequently
interpreted that way.
Where the two approaches lead to similar conclusions, we'll
generalize the discussion by referring to the ``upper
magnitude parameter."

Below we use the term $m_{\rm max}$ to signify the upper limit
of the magnitude variable in a likelihood map or the limit of
integration in an equation to compute a theoretical tectonic
moment arising from earthquakes.
As we will see below, in different approximations of
earthquake size distribution, $m_{\rm max}$ may have various
forms, though their estimated numerical values are usually
close.
We do not treat $m_{\rm max}$ as a firm limit, but we use it
as a convenient general reference because many other
researchers do so.

The statistical estimate of the maximum magnitude for global
earthquakes, including subduction zones and other tectonic
regions, yielded the values $m_{\rm max} \approx 8.3$ (Kagan
and Jackson, 2000).
The moment conservation provided an estimate for subduction
zones $m_{\rm max} = 8.5$ -- $8.7 \pm 0.3$ (Kagan, 1997;
2002b).
Moreover, the maximum earthquake size was shown to be the
indistinguishable, at least statistically, for all the
subduction zones studied.
Applying a combined statistical estimate and
moment conservation principle, Bird and Kagan (2004)
estimated the corner magnitude to be about 9.6.
As we explain below, the difference between the above
estimates (8.6 {\it vs} 9.6) is caused mainly by various
assumptions about the tectonic motion parameters.
These $m_{\rm max}$ determinations, combined with the
observation of very large ($m \ge 9.0$) events in the other
subduction zones (Stein and Okal, 2007, 2011; McCaffrey,
2007, 2008), should have warned of such a possible earthquake
in any major subduction zone, including the Sumatra and Tohoku
areas.

In Section~\ref{dist} below we consider two statistical
distributions for the earthquake scalar seismic moment
and statistical methods for evaluating their parameters.
Section~\ref{size} discusses the seismic moment conservation
principle and its implementation for determining the upper
magnitude parameter.
We then demonstrate how these techniques for size evaluation
work in subduction zones, showing that $m 9.0$ -- $m 9.7$
earthquakes can be expected in all major zones, including the
Tohoku area.
For the Tohoku area the approximate recurrence interval for $m
\ge 9.0$ earthquake is on the order of 350 years (Kagan and
Jackson, 2012, see more in the Discussion Section).
By the term `recurrence interval' we do not imply that large
earthquakes occur cyclically or quasi-periodically; contrary
to that we presented an evidence that all earthquakes
including the large ones are clustered in time and space
(Kagan and Jackson, 1999, see also the Discussion Section).

\section{Evaluation of earthquake size distribution for
subduction zones }
\label{dist}

\subsection{Earthquake catalogs
}
\label{dist1}

We studied earthquake distributions and clustering for the
global CMT catalog of moment tensor inversions compiled by
the GCMT group (Ekstr\"om {\it et al.}, 2005; Ekstr\"om, 2007;
Nettles {\it et al.}, 2011).
The present catalog contains more than 33,000 earthquake
entries for the period 1977/1/1 to 2010/12/31.
The event size is characterized by a scalar seismic
moment $M$.

We also analyzed the Centennial (1900-1999) catalog by Engdahl
and Villase\~nor (2002).
The catalog is complete down to magnitude 6.5 ($M_S$/$m_B$ or
their equivalent) during the period 1900-1963 and to 5.5
from 1964-1999.
Up to 8 different magnitudes for each earthquake are listed in
the catalog.
We use the maximum of the available magnitudes in the
Centennial catalog as a substitute for the moment magnitude
and construct a moment-frequency histogram.
There are 1623 shallow earthquakes in the catalog with $m
\ge 7.0$; of these 30 events have $m \ge 8.5$.

\subsection{Seismic moment/magnitude statistical distributions
}
\label{dist2}

In analyzing earthquakes here, we use the scalar seismic
moment $M$ directly, but for easy comparison and display we
convert it into an approximate moment magnitude using the
relationship (Hanks, 1992)
\be
m_W \ = \ {2 \over 3} \, (\, \log_{10}M - C \, ) \, ,
\label{Eq01}
\ee
where $C = 9.0$, if moment $M$ is measured in Newton m (Nm),
and $C = 16.0$ for moment $M$ expressed in {\sl dyne-cm} as in
the GCMT catalog.
The equation above provides a unique mapping from magnitude to
moment, so where appropriate we'll use both of the same
subscripts.
Thus $m_{\rm max}$ implies a corresponding $M_{\rm max}$, etc.

Since we are using the moment magnitude almost exclusively,
later we omit the subscript in $m_W$.
Unless specifically indicated, we use for consistency the
moment magnitude calculated as in (\ref{Eq01}) with the scalar
seismic moment from the GCMT catalog.

In this work we consider two statistical distributions of the
scalar seismic moment:
(a) the truncated Gutenberg-Richter (G-R) or equivalently
truncated Pareto distribution, in which the upper magnitude
parameter is the maximum magnitude, and
(b) the gamma distribution (Kagan, 2002a; 2002b), in which the
upper magnitude parameter is the corner magnitude $m_{\rm
cg}$.

For the truncated Pareto distribution the probability density
(pdf) is
\be
\phi (M) \ = \ { { M_{xp}^\beta M_t^\beta }
\over { M_{xp}^\beta - M_t^\beta } } \beta M^{ -1 - \beta}
\quad {\rm for} \quad M_t \le M \le M_{xp} \, .
\label{Eq02}
\ee
Here $M_{xp}$ is the upper truncation parameter,
$M_t$ is the lower moment threshold (the smallest moment above
which the catalogue can be considered to be complete),
$\beta$ is the index parameter of the distribution.
Note that $ \, \beta = {2 \over 3} b \, $, where $b$ is the
familiar $b$-value of the Gutenberg-Richter distribution
(Gutenberg and Richter, 1954, pp.~16-25).

The gamma distribution has the pdf
\ba
\phi (M) & = C^{-1} {\beta \over M} (M_t/M)^\beta
\, \exp [ (M_t - M )/M_{cg} ],
\nonumber\\
& \quad {\rm for} \quad M_t \le M < \infty,
\label{Eq03}
\ea
where $M_{cg}$ is the corner moment parameter controlling the
distribution in the upper ranges of $M$ (`the corner moment')
and $C$ is a normalizing coefficient.
Specifically,
\begin{equation}
C \ = \ 1 \ - \ (M_t/M_{cg})^{\beta}
\exp (M_t/M_{cg}) \ \Gamma (1 - \beta, M_t/M_{cg}) \, ,
\label{Eq04}
\end{equation}
where $\Gamma$ is the gamma function (Bateman and Erdelyi,
1953). For $M_{cg} > > M_t$ the coefficient $C \approx 1$.
Below we simplify the notation $M_{px}$ and $M_{cg}$ as $M_x$
and $M_c$, respectively, keeping the notation $M_{\rm max}$ to
represent either $M_{px}$ or $M_{cg}$ as appropriate.
Thus, each distribution is controlled by two parameters:
its slope for small and moderate earthquakes, $\beta$, and
its maximum or corner moment $M_x$ or $M_c$ describing
the behavior of the largest earthquakes.
The two equations above are normalized distributions.
Both need a multiplicative constant, the threshold earthquake
rate, to calculate rates at any magnitude above the threshold.

In Fig.~\ref{fig01} we show the magnitude-frequency curves for
shallow earthquakes (depth less or equal to 70~km) in
the Japan-Kurile-Kamchatka \#19 Flinn-Engdahl (Flinn {\it et
al.}, 1974; Young {\it et al.}, 1996) zone for the period
1977-2010 (i.e., before the Tohoku event).
Similar curves for period before and after 2011/1/1 for the
rupture zone of the Tohoku earthquake are shown in our
companion paper (Kagan and Jackson, 2012, Figs.~1,~2).
Knowing the distribution of the seismic moment, one can
calculate occurrence rates for earthquakes of any size, so we
need a reliable statistical technique to determine the
parameters of a distribution.

\subsection{Likelihood evaluation of distribution parameters }
\label{dist3}

We applied the likelihood method to obtain estimates of
$\beta$ and $m_c$ (Kagan, 1997, 2002a).
Fig.~\ref{fig02} displays the map of the log-likelihood
function for two parameters of the gamma distribution.
The $\beta$-value (around $0.61 \pm 0.038$) can be determined
from the plot with sufficient accuracy (Fig.~\ref{fig02}), but
the corner magnitude evaluation encounters serious
difficulties.
The upper contour of the 95\% confidence area in the
likelihood map is not well-constrained and allows infinite
$M_x$.
This means that only the lower limit for $m_c$, around 8.2,
can be reliably evaluated with the available data. Hence the
maximum likelihood estimate ($m_c = 8.7$) is not
well-constrained by the likelihood map.
However, even the lower limit for $m_c$ is higher than the
maximum magnitude size ($m7$ to $m8$) proposed by the official
hazard map for the Tohoku area (Headquarters for Earthquake
Research Promotion, 2005; Seismic Activity in Japan, 2008;
Simons {\it et al.}, 2011).
Thus, a simple statistical test could have revealed that
earthquakes much larger than magnitude 7 to 8 should be
expected.

Fig.~\ref{fig02} is similar to the likelihood map constructed
by Bird and Kagan (2004, Fig.~7F; see also Table~5) for all
subduction zones taken together.
The calculation in that plot was made with the tapered
G-R distribution (TGR) discussed by Kagan (2002a).
That distribution is similar in spirit to the gamma
distribution, except that the taper is applied to the
cumulative rather than the density distribution.
For $\beta = 0.64$ and identical total moment rates, the
$m_c$-values of the TGR are 0.39 lower than those for the
gamma distribution (Kagan, 2002b; see also Kagan and Jackson,
2000, Fig.~2 and its discussion).

From Bird and Kagan's map (2004, Fig.~7F), the lower limit for
$m_{cm}$ is about 9.0, but even the complete 20-th century
earthquake record is insufficient to obtain the upper
statistical limit.
Kagan (1997) has showed that all subduction zones have
essentially the same maximum magnitude parameters;
this result implies that the corner magnitude $m_{cg}$
is at least 9.4 in the major zones.

Geller (2011) and Stein {\it et al.}\ (2011) suggest that a
major reason for grossly under-estimating the maximum
magnitude for Japanese earthquakes is that many seismologists
accept the flawed seismic gap model based on the
characteristic earthquake hypothesis (see, for example,
Wesnousky {\it et al.}, 1984; Annaka {\it et al.}, 2007, and
the Introduction Section).
This inadequate model suggests that a fault can be subdivided
into segments and a maximum allowable event on such a segment
is limited either by its length or by the available
historic or instrumental record.
Bird (2010) found that even ``diligent and extensive mapping
of faults [cannot] provide reliable estimates of the expected
maximum earthquakes at these faults".
Kagan and Jackson (1991) reported serious problems with the
seismic gap model.
Jackson and Kagan (2011) summarize theoretical and
observational arguments against the seismic gap/characteristic
earthquake model.
Simons {\it et al.}\ (2011, p. 1425) also present some
evidence contradicting this hypothesis, indicating that
historical events do not repeat one another and their slip may
change significantly.
They conclude ``the concept of a characteristic subduction
earthquake with approximately the same slip per event at a
given location may be of limited use".

\section{Seismic moment conservation principle }
\label{size}

We try to estimate the upper bound of the seismic
moment-frequency relation, using the moment conservation
principle as another, more effective method for determining
the maximum/corner magnitude.
Quantitative plate tectonics and space geodetic methods
currently provide a numerical estimate of the tectonic
deformation rate for all major tectonic plate boundaries and
continental regions of significant distributed deformation
(Bird and Kagan, 2004; Kagan {\it et al.}, 2010).
We compare these estimates with a similar one for the
seismic moment release.

The seismic moment rate depends on three variables (see
Eqs.~\ref{Eq06a}, \ref{Eq06b} below) --
\hfil\break
$\bullet$ \quad \
1. \ The number of earthquakes in a region ($N$),
\hfil\break
$\bullet$ \quad \
2. \ The $\beta$-value (or $b$-value) of the G-R relation,
\hfil\break
$\bullet$ \quad \
3. \ The value of the maximum (corner) magnitude $m_c$.

The tectonic moment rate $\dot M_T$ depends on the following
three variables which are not well-known --
\hfil\break
$\bullet$ \quad \
1. \ The seismogenic zone width ($W$ -- 30-104 km),
\hfil\break
$\bullet$ \quad \
2. \ The seismic efficiency (coupling) coefficient ($\chi$ --
50-100\%),
\hfil\break
$\bullet$ \quad
3. \ The value of the shear modulus ($\mu$ -- 30-49GPa).
\be
\dot M_T \ = \ \chi \, \mu \, W \, {\cal L} \, \dot u \,
,
\label{Eq05}
\ee
where $\dot u$ is the slip rate, ${\cal L}$ is the length of
a fault (compare Eq.~13 by Kagan, 2002b).

\subsection{Area-specific conservation principle }
\label{size1}

The discussion in this Subsection is based generally on our
previous papers (Kagan, 2002a; 2002b).
In those papers we consider two theoretical moment-frequency
models:
(a) a truncated Pareto distribution;
\be
\dot M_s \ = \
\ { { \alpha_0 \, M_0^\beta \, \beta } \over {1 -
\beta }} \, M_{x}^{1 - \beta} \, \xi_{p}
\, ,
\label{Eq06a}
\ee
where $\dot M_s$ is the seismic moment rate and $\alpha_0$ is
the rate of occurrence for events with moments larger or equal
$M_0$; $\alpha_0 = N/\Delta T$ with $ \Delta T$ as the catalog
duration.
In most cases $M_0$ can be chosen to correspond to the
observational threshold moment $M_t$.
Coefficient $\xi$ is a correction factor needed if the
distribution is left-truncated close to the maximum or the
corner moment; under usual circumstances it equals 1.0.
For the gamma distribution (b) the analogous formula is
\be
\dot M_s \ = \
{ { \alpha_0 \, M_0^\beta \, \beta } \over {1 -
\beta }} \, M_{c}^{1 - \beta} \, \Gamma (2 - \beta) \,
\xi_{g} \, .
\label{Eq06b}
\ee

We assume that the two theoretical laws (Eqs.~\ref{Eq06a} and
\ref{Eq06b}) describe a distribution with the same moment rate
$\dot M_s$ and seismic rate of occurrence
$\alpha_0$.
Using Eqs.~\ref{Eq06a} and \ref{Eq06b}, relations between the
maximum or corner moments can be specified
\be
M_{xp} \, \beta^{1/(1 - \beta)} \ = \
M_{cg} \, [\beta \, \Gamma(2 - \beta)]^{1/(1 - \beta)} \, .
\label{Eq07}
\ee
The gamma function $\Gamma(2 - \beta)$ changes slowly in the
range of $\beta$-values encountered in the moment-frequency
relations: for values of $\beta$ in the interval 1/2 to 2/3,
the gamma function is $\Gamma(2-1/2) = 0.886$ or
$\Gamma(2-2/3) = 0.893$.
Therefore, the difference between the maximum and corner
magnitudes is relatively small: for southern California our
calculations (Kagan, 2002b, Fig.~2) yield the magnitude values
$m_x = 8.35$, and $m_c = 8.45$ for the two distributions shown
in Eqs.~\ref{Eq02} and \ref{Eq03}.

Fig.~\ref{fig03} shows the $\beta$-values determined for 18
Flinn-Engdahl (FE) zones (Gutenberg and Richter, 1954, Fig.~1;
Flinn {\it et al.}, 1974; Young {\it et al.}, 1996) listed
in sequential order.
These FE regions correspond to major subduction zones and
they have been selected by us because the FE regionalization
had been defined before the GCMT catalog started, thus
eliminating selection bias.
It is also easier to replicate our results (the programs and
tables for the Flinn-Engdahl zones are publicly available, see
Section `Data and Resources').
In this plot we use the GCMT catalog at the same temporal
interval as in our previous paper (Kagan, 1997).
In Fig.~\ref{fig04} the catalog duration is extended to
the end of 2010.
Both plots demonstrate that (a) the $\beta$-values do not
depend significantly on the catalog duration, though their
standard errors do decrease with the duration and earthquake
numbers increase;
(b) the $\beta$-values are approximately the same for all the
zones, and the hypothesis of the values equality cannot be
statistically rejected (Kagan, 1997).
The additional data since 1995 makes the argument for a common
$\beta$ much stronger.

Figs.~\ref{fig05}, \ref{fig06}, and \ref{fig07} show the
distribution of the corner magnitude obtained, using
Eq.~\ref{Eq06b}, for the earthquake catalogs of different time
duration.
Estimates of $m_c$ for all diagrams in all the subduction
zones are statistically indistinguishable.
This means that all such zones should have the same
maximum or corner magnitude.

The estimate of $m_c$ depends on the parameter value used to
calculate the tectonic moment rate (Eq.~\ref{Eq05}): for
$\beta = 2/3$, the change of any parameters (such as $W$ or
$\chi$) by a factor of two implies an increase or decrease of
the $m_c$ by about 0.6 (Kagan, 2002b, Eq.~17).
We see this influence by comparing Table~\ref{Table1} with
results for the subduction zones in similar Table~1 by Kagan
(1997), where the parameters used for calculating tectonic
rate were $W$~=~30~km, $\mu~=~30$~GPa, $\chi~=~1.0$.
The difference in the $m_c$ estimates for the two tables is
caused mainly by changes in the above parameters.

We also compare the $m_c$-values for the same zones in
Figs.~\ref{fig05}, \ref{fig06}, and \ref{fig07}.
The values in Figs.~\ref{fig06} and \ref{fig07} differ
greatly: by 0.5 and more in five zones: Kermadec, Fiji,
Japan-Ryukyu, Sunda, and Andaman (FE12, 13, 20, 24, 46).
This is not surprising since the magnitudes in the Centennial
catalog were determined with large random and systematic
errors.
In addition, given the high magnitude threshold, the
earthquake numbers will have significant random fluctuations.
When matching up Figs.~\ref{fig05} and \ref{fig06}, only one
zone, Andaman-Sumatra, shows a $m_c$ difference of about
0.9.
This finding is due to the 2004 Sumatra earthquake and its
aftershocks which significantly increased the total number of
events to 143 in 34 years (Table~\ref{Table1}) {\it versus} 22
earthquakes in 19.5 years in Table~1 by Kagan (1997).
The annual rate increase is by a factor 3.15, which for $\beta
= 2/3$ corresponds to $m_c$ decrease of 1.0.
In our $m_c$ calculations (Eq.~\ref{Eq06b}) we use the
$\beta$-values determined for each zone.

As mentioned, the hypothesis that the $m_c$-values are the
same within their uncertainties for all subduction zones
considered cannot be rejected with statistical significance.
Thus, the conjecture that $m_c \approx 9.0-9.7$ in all such
zones is supported by comparing the theoretical estimates with
measured magnitude values in several subduction zones.
For example, a $m9.0$ earthquake occurred in zone \#19
(Kamchatka, Russia) on November 4, 1952, confirming that this
subduction zone could experience large earthquakes beyond
$m9$.

In Table~\ref{Table1}) we also calculate parameter values for
the Tohoku area (latitudes $35-40^\circ$N, longitudes
$140-146^\circ$E).
The length of this trench zone is around 620~km in this
spherical rectangle.
The value of the tectonic moment accumulation rate attributed
to the subducting Pacific plate is compatible with that
proposed by Ozawa {\it et al.} (2011) in the Japan trench area
from latitude $36^\circ$N to $39.5^\circ$N ($1.63 \times
10^{20}$ Nm/y).
Ozawa {\it et al.} (2011) also suggest using $\mu = 40$ GPa
for the Japan trench.

Whereas the values of the parameters $\beta$ and $m_c$ are
approximately the same for the Tohoku area and Flinn-Engdahl
zone \#19, other entries for these rows (12 and 12a) differ
significantly. This means that the maximum observed earthquake
or the ratio of the seismic rate to tectonic rates ($\psi$)
for individual subduction zones varies greatly and cannot be
used reliably to characterize area seismicity.

For about 110 years of the instrumental seismic record,
five zones have experienced earthquakes with a magnitude~9
or larger.
Figs.~\ref{fig05}--\ref{fig07} also show that for the longer
catalog, the average maximum observed magnitude approaches the
average estimate of $m_c$.
This suggests that if the available earthquake record duration
were comparable to the recurrence time of the largest
earthquakes (a few hundred years), the difference between the
observed maximum magnitude ($m_o$) and $m_c$ would largely
disappear.

Using the parameter values for the moment-frequency
distribution determined by Bird and Kagan (2004, Table~5) for
all the subduction zones ($b=0.96$, $M_t=3.5\times
10^{17}$~Nm, $\alpha_t = 76.74$~eq/y, $m_c=9.58$), we
calculate the number of $m \ge 9$ events expected to occur
worldwide over a century
\be
N (m>9) \ = \ 100.0 \times \alpha_t \times 10^{-0.96 \,
(9.0 \, - \, 5.696)} \ = \ 5.2 \, .
\label{Eq08}
\ee
In fact five large earthquakes with magnitude~9 or greater did
occur in the last 100 years (see
Figs.~\ref{fig05}--\ref{fig07}).
The distribution parameters (Bird and Kagan, 2004) were
estimated before two recent giant earthquakes struck, so the
almost perfect correspondence can be considered a coincidence.

Fig.~\ref{fig08} again demonstrates how the catalog duration
affects the ratio of the seismic rate to the tectonic rate for
different subduction zones.
That ratio is below one for a shorter catalog, but increases
to a value close to 1.0 for a longer list.
This increase is caused mainly by a few large earthquakes that
struck South America and Sumatra regions.
As Zaliapin {\it et al.}\ (2005) show, a sum of the scalar
earthquake moments varies widely due to their power-law
distribution.

\subsection{Geometric self-similarity of earthquake rupture
}
\label{size2}

Fig.~\ref{fig09} displays an update of Fig.~6a by Kagan
(2002c).
Several recent mega-earthquakes are included in the plot: the
2004 Sumatra ($m9.1$) and the Tohoku ($m9.2$) events -- the
right-hand symbols in the diagram, as well as preliminary
results for two 2012/04/11 strike-slip earthquakes off the
Sumatra coast (McGuire and Beroza, 2012).
As mentioned below Eq.~\ref{Eq01}, for all earthquake
magnitudes we apply (\ref{Eq01}) to the available GCMT scalar
moment.
Despite the differences in the aftershock zone lengths for the
Sumatra and the Tohoku events, emphasized in many
publications, the symbols for these earthquakes on the graph
are not far away.
The vertical difference is smaller than the scatter for
moderate earthquakes.
The seeming contradiction of their size evaluation is caused
by various techniques employed in measuring the rupture size.
We use the same measurement method for all earthquakes:
namely, a fit of the aftershock spatial scatter by a
two-dimensional Gaussian distribution (Kagan, 2002c).
For Sumatra and Tohoku events we obtain the aftershock zone
size ($2 \times \sigma$ confidence area length) 905 and
533~km, respectively.
Two major strike-slip events off the Sumatra coast ($m8.6$ and
$m8.3$, shown by `$+$' symbols in the plot) are only slightly
above the regression line for all earthquakes, thus they
follow a common relation between the earthquake size and the
length of aftershock zone.

Comparing the regression results from Fig.~\ref{fig09} with
those of Fig.~6a by Kagan (2002c) demonstrates that the
scaling parameter estimates are robust.
Since 2000 these earthquake numbers have increased by almost a
factor of two.
Moreover, in Fig.~\ref{fig09} there are three major ($m \ge
8.8$) events, whereas the largest earthquake in the 2002 study
was $m8.4$.
However, the values of regression coefficients in both
datasets are essentially the same.

Fig.~\ref{fig09} shows both linear and quadratic regression
curves for log length versus magnitude.
There is practically no difference between linear and
quadratic fits.
No observable scaling break or saturation occurs for the
largest earthquakes of different focal mechanisms; thus, the
earthquake geometrical focal zones are self-similar.
Assuming self-similarity, we adopted the following scaling for
the average length ($L$), the average downdip width ($W$), and
the average slip ($U$) as a function of the moment ($M$):
the length ($L$) is proportional to the cube root of the
moment: $L \propto \sqrt [3] M$, implying a self-similarity of
the earthquake rupture pattern, i.e., $W$ and $U$ are also
proportional to $ \sqrt [3] M$.
The average slip is proportional to the average length
implying nearly constant value for stress/strain drop which is
related to $U/L$.

The above results imply that the earthquake slip penetrates
well below the seismogenic layer during large earthquakes.
Shaw and Wesnousky (2008) and McGuire and Beroza (2012) also
note that significant coseismic slip occurs below the
seismogenic layer.
If the downdip seismogenic width $W$ changes for the largest
earthquakes, it may influence the calculation of tectonic rate
$\dot M_T$ (see Eq.~\ref{Eq05}).

\subsection{Conservation principle for faults }
\label{size3}

Here we consider moment conservation in the specific case that
tectonic deformations are dominated by a fault or plate
boundary with estimated slip rate.
For simplicity we'll refer to plate boundary surfaces as
faults.

Many attempts have been made to compare the slip budget at
a subduction zone with its release by earthquakes.
McCaffrey (2007, 2008) compared the slip values at the global
subduction zones with their release by earthquakes spanning
the whole length of a zone.
He found that for practically for all the zones, $m9$ and
greater earthquakes are possible with recurrence times on the
order of a few centuries.

In particular, for Japan McCaffrey (2008) calculated the
maximum moment $M_{\rm max}$ as
\be
M_{\rm max} \ = \ \mu \, \overline u \, L \, Z_{\rm max} \,
/ \, \sin \delta \, ,
\label{Eq09}
\ee
where $\overline u$ is the average slip, $Z_{\rm max}$ is the
maximum depth of the slip (40~km used), and $\delta$ is the
average fault dip angle (taken to be $22^\circ$ for Japan),
implying $W = 106.8$~km.
The recurrence time for the maximum earthquake is
\be
T \ = \ \overline u \, / \, f \, \chi \, \nu \, ,
\label{Eq10}
\ee
where $\nu$ is the plate motion rate, $f$ is the fraction
of the total seismic moment in $m9$ earthquakes, and
$ \overline u = 2.5 \times 10^{-5} L$.
The parameter $f$ is taken to be equal to $1-\beta$
(apparently using results for the characteristic earthquake
distribution by Kagan, 2002a; 2002b).
By taking $L~=~654$~km, $\beta=0.57$, $\nu~=~62-81$~mm/y,
and $\chi=1$, McCaffrey obtains $M_{\rm max} = 10^{22.53}$~Nm,
i.e., $m_{\rm max} = 9.0$, and $T \approx 532$~y.

Simons {\it et al.}\ (2011) considered how the slip in the
Tohoku area on the order $\nu~=~80-85$~mm/y is accommodated by
subduction earthquakes.
They proposed that only very large events, similar to the
Tohoku $m9.2$ earthquake, can explain this displacement
rate.
Simons {\it et al.}\ (2011) included only the largest
earthquakes in the slip budget.
However, events smaller than the maximum earthquake also
contribute to the slip budget, and all earthquakes need to be
considered in boundary-specific calculations.

A discussion in this Subsection is based broadly on one of our
previous papers (Kagan, 2005).
Several issues need to be noted in fault-specific slip
calculations:
\hfil\break
$\bullet$ \quad \
1. \ The form of the general (area-specific) distribution of
earthquake sizes.
To simplify calculations we take it as the truncated Pareto
distribution (see Eq.~\ref{Eq02}).
\hfil\break
$\bullet$ \quad \
2. \ The fault-specific moment distribution -- large
earthquakes have a bigger chance to intersect a surface; hence
the moment distribution differs from area-specific concerns.
\hfil\break
$\bullet$ \quad \
3. \ The geometric scaling of earthquake rupture.
As described earlier, length-width-slips are scale-invariant,
i.e., for an earthquake of magnitude $m$: $L_m, W_m, u_m
\propto
\root 3 \of M$.
\hfil\break
$\bullet$ \quad \
4. \ Geometric self-similarity of earthquake rupture implies
that the earthquake depth distribution would differ for small
versus large shocks: at least for strike-slip earthquakes,
large events would penetrate below the seismogenic layer.
For thrust and normal events, the consequences of geometric
self-similarity are not clear; their depth distribution has
not been sufficiently studied.
\hfil\break
$\bullet$ \quad \
5. \ Most small earthquakes do not reach the Earth's
surface and therefore do not contribute to surface fault
slip.
The contribution of small earthquakes needs to be properly
computed.

Because the distribution of slip with depth is poorly
understood, we calculate the maximum earthquake size for
several special simple cases.
An earthquake of moment $M$ (magnitude $m$) is specified as
\be
M \ = \ \mu \, L_m \, W_m \, \overline u_m \, .
\label{Eq11}
\ee
Using the results shown in Fig.~\ref{fig09}, we presume for an
earthquake of magnitude $m=7.0$ or moment $M = 10^{19.5}$~Nm,
that $L_7 = 60$~km, $W_7=10$~km, and $\overline u_7 = 1.76$~m
or $\overline u_7 = 1.076$~m, depending on the value of the
shear modulus: $\mu = 30$~GPa or $\mu = 49$~GPa, respectively.

Slip distribution over the fault plane is highly
non-uniform in a horizontal direction (Manighetti {\it et
al.}, 2005, 2009).
Kagan (2005, [51]) argued that `the slip of large earthquakes
should ``catch up" with the slip deficit at the Earth's
surface left by smaller events'.
Thus, the slip of large events must be larger at the surface
than in the middle of a seismogenic zone.
Fialko {\it et al.}\ (2005) and Kaneko and Fialko (2011) show
several examples of $m 7$ earthquakes which exhibit a strong
slip deficit close to the Earth's surface.
This may imply that the seismic efficiency coefficient
($\chi$) may also change with depth.
However, since we lack reliable data, for our approximate
calculations, we take slip to be uniform over a rectangle $L_m
\times W_m$.

We specify earthquake magnitude $m$ rupture dimensions as
\be
L_m \ = \ L_7 \times \Bigl ( \, 10^{1.5 \, m + 9} /10^{19.5}
\,
\Bigr ) \, ^{1/3} \, .
\label{Eq12}
\ee
Analogous expressions are used for $W_m$ and $\overline u_m$.

The relationship between surface slip and the maximum
earthquake size depends on the extent to which earthquakes of
a given size break the surface.
We consider below three special cases.
In the first, we assume earthquakes of all sizes are uniformly
distributed on the fault surface.
In the second, we assume all earthquakes larger than a given
magnitude, and only those, break the surface.
In the third, we assume that small earthquakes are uniformly
distributed above a given depth, but larger ones may penetrate
deeper.

In the first simple case of the maximum earthquake size
calculation, we suppose that earthquakes of all sizes are
distributed uniformly over a fault surface of width $W$.
Then, as we mentioned above, for an earthquake with magnitude
$m$, the surface slip contribution $u_m^\prime$ would be
\be
u_m^\prime \ = \ u_m \times W_m / W \, \quad {\rm for} \quad \
W \ge W_m \, ,
\label{Eq13}
\ee
accounting for the fact that only a few of the smaller
earthquakes would reach the surface.

In deriving formulas for boundary-specific surface displacement
$U_s$, we simplify Kagan's (2005) results, taking into account
the self-similarity of earthquake rupture (\ref{Eq12})
\be
U_s \ = \ { { \lambda_m \, u_m ^\prime \, L_m } \over
{ 1.5 - b} }
\, ,
\label{Eq14}
\ee
where $\lambda_m = \alpha_m/\cal L$ is the rate of earthquakes
with magnitude greater or equal $m$ per km of a fault surface.
If $W_m = W$ is due to assumed scaling, the resulting $U_s$
would always be the same, since an increase in $u_m$ and $L_m$
would be compensated by a decrease in $\lambda_m$.

Making some order-of-magnitude calculations, we estimate the
slip rate for the Tohoku area.
If the magnitude is not close to the maximum, $\alpha_m$
scales with magnitude $m$ as $10^{\, b \, (7-m)}$ (see
Eq.~\ref{Eq02}).
Thus, from Fig.~\ref{fig01} the annual rate of $m5.8$
earthquakes is 425/34, and $\beta = 0.61$.
Hence $\alpha_7 \ = \ (425/34) \times 10^{- 0.61 \times 1.5
\times 1.2} \ = \ 0.997 \ \simeq \ 1.0$ and $\cal L \ = \
$~3,000~km.
Then for $b=1$ and $m=7$ we obtain $U_s = 7$~mm/y.
For $m=8$ the slip rate is $U_s = 22$~mm/y, and $U_s =
70$~mm/y for $m=9$.
If we replace $\mu = 30$~GPa by $\mu = 49$~GPa, then for
$m=9$ the slip is $U_s = 43$~mm/y.
All these values are smaller than $U_s = 82.5$~mm/y suggested
by Simons {\it et al.}\ (2011) for the Tohoku area.
Therefore, the results do not support this case of fault
surface slip distribution, unless the present rate of $m5.8$
and larger earthquakes is below the long term average.

For the second case we presume that all surface slip is due to
earthquakes exceeding a certain size ($m_f$) which always
rupture the Earth's surface.
In effect, we suppose that small earthquakes do not contribute
to the surface slip.
Such a model may be appropriate for strike-skip faults like
those in California, where very few small events occur near
the surface (Kagan, 2005).
It is possible that even for subduction zones, this model
would produce more correct results.

Then for $b \ne 1.0$ we obtain
\be
U_s \ = \ { { \lambda_f \, u_f \, L_f } \over
{ 1.0 - b} }
\times \Bigr [10^{(b-1.0) (m_x - m_f)} - 1 \Bigr ]
\, ,
\label{Eq15}
\ee
and for $b=1$
\be
U_s \ = \ \lambda_f \, u_f \, L_f \, \log 10 \times (m_x -
m_f)
\, .
\label{Eq16}
\ee

Several calculations can be made with these formulas to get a
rough estimate of the maximum magnitude $m_x$ needed to obtain
the slip rate $U_s = 82.5$~mm/y.
For $b=1$ and $\mu = 30$~GPa, $m_x = m_f + 1.02$.
Therefore, if we assume $m_f = 7.0$, $m_x = 8.02$ and for $m_f
= 8.0$, as more appropriate for subduction zones, $m_x =
9.02$.
These values depend, of course, on the presumed parameters of
the earthquake rupture.
If we take the values $ L_7 = 37.5$~km, $W_7 = 15$~km, and
$\overline u_7 = 1.87$~m, as suggested by Kagan (2005), the
estimate of the maximum magnitude changes to $m_x = m_f +
1.53$.
A similar increase of $m_x$ occurs if we modify $\mu$: for
$\mu = 49$~GPa, $m_x = m_f + 1.65$.
If we change the $b$-value, for instance, take $b=0.9$, but
keep $\mu = 30$~GPa, then $m_x = m_f + 0.91$.

For the third case, we consider a combination of two models
(Eqs.~\ref{Eq14} and \ref{Eq16}): we suppose that in the upper
part of the fault surface with the width $W_f$, small
earthquakes are distributed uniformly, whereas large
earthquakes $m \ge m_f$ penetrate deeper.
Thus, the total slip would be a sum of two terms: one
reflecting the contribution of small and moderate events and
the other from large earthquakes.
Then taking $b=1$, $m_f = 8$, $W_f = 31.6$~km, and $\mu =
49$~GPa, we obtain $m_x = 8.79$.

Calculations in this Section are more subjective than those in
the previous Subsection~\ref{size1}; unfortunately the
distributions of the slip and the earthquake depth have not
yet been studied as thoroughly as the area-specific
magnitude-frequency relation.
However, these approximate computations imply that the maximum
magnitude in the Flinn-Engdahl zone \#19 and in the Tohoku
area is around 9.0, i.e., much greater than was assumed in the
various hazard maps for Japan compiled by many investigators
(see the Introduction).

\section{Discussion }
\label{disc}

It is commonly believed that after a large earthquake, its
focal area ``has been de-stressed" (see, for example, Matthews
{\it et al.}, 2002), thus lowering the probability of a new
large event in this place, though it can increase in nearby
zones.
This reasoning goes back to the flawed seismic
gap/characteristic earthquake model (Jackson and Kagan, 2011).
Kagan and Jackson (1999) showed that earthquakes as large as
7.5 and larger often occur in practically the same area soon
after a previous event.
Michael (2011) shows that earthquakes as large as $m8.5$ are
clustered in time and space: thus, such a big event does not
protect its focal area from the next giant shock.

Any forecast scheme that extrapolates the past instrumental
seismicity record would predict future moderate earthquakes
reasonably well.
However, as the history of the Tohoku area shows, we need
a different tool to forecast the largest possible events.
In our forecasts we consider the earthquake rate to be
independent of the earthquake size distribution, so the latter
needs to be specified separately.

Why is it so difficult to determine the maximum earthquake
size for each subduction zone and its recurrence period?
This question is especially important after two unpredicted
giant earthquakes: the 2004 Sumatra and the 2011 Tohoku.
Our available earthquake record is so short that it is
difficult to obtain this information by simple observation.

As indicated earlier, the seismic moment conservation
principle can answer our questions.
The general idea of moment conservation was suggested
some time ago (Brune, 1968; Wyss, 1973; Molnar, 1979;
Anderson, 1979).
However, without knowledge of the earthquake size
distribution, calculating the rate of huge earthquakes leads
to uncertain or contradictory results.
The classical G-R relation is not helpful in this respect
because it implies infinitely large earthquakes.
Only a modification of the G-R law that limits the upper
moment can provide a tool to match earthquake and moment
rates.
Kagan and Jackson (2000) and Kagan (2002a, 2002b) propose such
distributions defined by two parameters, $\beta$ and variants
of $m_{\rm max}$.

Applying these distributions also allows us to address
the problem of evaluating the recurrence period for these
large earthquakes.
Determining maximum earthquake size either by
historical or instrumental observations or by qualitative
analogies does not provide such an estimate: a similar
earthquake may occur hundreds or tens of thousand years later.
But Fig.~\ref{fig01} shows how using statistical distributions
of earthquake magnitudes may facilitate such calculations.

As we discussed in Subsections~\ref{size1} and \ref{size3},
the moment conservation principle allows us to
determine the upper magnitude parameter quantitatively.
In this respect area-specific calculations provide a more
precise size evaluation for many tectonic zones and, most
importantly, show that the subduction zones effectively
share the same upper magnitude parameter (Kagan, 1997).
Corner magnitude estimates based on moment conservation are
still annoyingly imprecise, with estimates in the range of 8.5
to 9.7.
The fact that several subduction zones have been hit with
giant earthquakes in the last 110 years suggests that all such
zones could experience earthquakes with the corner magnitude
towards the top end of that range.

Boundary-specific calculations are not yet as accurate and
reliable as the area-specific, and the computation for several
subduction zones has not been performed.
However, even the approximate estimates in
Subsection~\ref{size3} suggest that $m \ge 9$ is an
appropriate earthquake size for the Tohoku area.

The seismic moment-frequency relation (Fig.~\ref{fig01} and
Table~\ref{Table1}) which is based on earthquake statistics
for moderate events and the estimate of the corner moment by
the moment conservation principle, implies that the return time
of magnitude 9 or larger earthquakes off Tohoku is in the
range 300 to 400 years.
Uchida and Matsuzawa (2011) suggest a recurrence interval
of 260-880 years for the $m9$ events.
Simons {\it et al.}\ (2011) propose a 500-1,000 year interval
based in part on the historic record.

Is our estimated recurrence time of 300 to 400 years, based on
instrumentally recorded earthquakes and tectonic strain rates,
consistent with the historic record of magnitude 9 and larger
events?
In one interpretation, no such earthquake occurred off
Tohoku between the Jogan earthquake and tsunami in year 869
(Minoura {\it et al.}, 2001) and the 2011 Tohoku event.
That implies an empty interval of 1142 years.
Assuming Poisson recurrence, the probability of an interval
that long or longer is 5\% for a recurrence time of 382 years.
A longer recurrence time could not be rejected with 95\%
confidence based on the empty interval.
Minoura {\it et al.}\ suggested that there may have been three
Tohoku-sized tsunamis in the last 3,000 years before 2001.
Again assuming Poisson recurrence, there is a 5\% chance of
observing three or fewer events in 3,000 years for a recurrence
time of 387 years.
Longer recurrence times cannot be rejected with 95\%
confidence.
The above calculations assume that the observations are
complete for 1,142 and 3,000 years, respectively.
The sizes of the reported earthquakes are based on tsunami
run-up, and it is quite possible for larger earthquakes to
have smaller tsunami.
For example, Koketsu and Yokota (2011) suggest that the 1611
Keicho earthquake may have been about as large as the 2011
event.
Thus the historic record, based primarily on tsunami data,
does not conflict with our recurrence time estimate of 300 --
400 years,

In conclusion, we would like to evaluate the upper magnitude
limit for the subduction zones as well as recurrence intervals
for such earthquakes.
Two upper global estimates can be calculated: for the gamma
distribution, we take the values from Table~\ref{Table1}
$m_{cg} = 9.36\pm0.27$ to get the 95\% upper limit $m_{cg} =
9.9$.
Bird and Kagan (2004, Table~5) determined for the tapered G-R
(TGR) distribution $m_{cm} = 9.58^{+\infty}_{-0.23}$, and
the approximate 95\% upper limit $m_{cm} = 10.1$.
For the sake of simplicity, we take $m_{\rm max} = 10.0$.
Calculations similar to (\ref{Eq08}) can be made to obtain an
approximate estimate for the average inter-earthquake period.
From Fig.~1b by Kagan (2002a) one can determine the return
period as it differs from the regular G-R law: the gamma
distribution cumulative function at $m_c$ is below the G-R
line by a factor of about 10.
For the TGR distribution the factor is $e$.
Thus, for the gamma distribution, the recurrence time for the
global occurrence of the $m \ge 10.0$ earthquake is about
1,750~years; for the TGR distribution this period is about
475~years.
Of course, the distributions in these calculations are
extrapolated beyond the limit of their parameters' evaluation
range.
But the above recurrence periods provide a rough idea how big
such earthquakes could be and how frequently they might occur
worldwide.

According to the same reasoning for the Flinn-Engdahl \#19
zone a $m \ge 10.0$ earthquake could repeat in about 9,000 or
32,000 years for the TGR and the gamma distributions,
respectively.
The rupture length of the $m10$ event can be estimated from
Fig.~\ref{fig09}: at about 2,100~km it is comparable to
the 3,000~km length of zone \#19.
These long recurrence periods indicate that it would be
difficult to find displacement traces or the tsunami record
for these earthquakes in the paleo-seismic or paleo-tsunami
investigations which usually extend over the period of a few
thousand years (Wesnousky {\it et al.}, 1984; Nishenko, 1991;
Minoura {\it et al.}, 2001; Grunewald and Stein, 2006).
Moreover, since subduction zone faults are mostly offshore,
finding their displacement trace is impractical.

\section{Conclusions }
\label{conc}
A magnitude 9 earthquake off Tohoku should not have been a
surprise.
Since 1997 there has been evidence that subduction zones have
indistinguishable $b$-values and upper magnitude parameters.
Four previous subduction zone earthquakes of magnitude 9 or
larger around the globe in the last 110 years should have
served as warning.
Under reasonable assumptions of plate boundary properties,
earthquakes at least as large as 8.5 are required to satisfy
the geodetically and geologically observed moment rates.
The seismic gap/characteristic earthquakes model, which formed
the basis for smaller magnitude limits, was shown to be
inadequate as early as 1991.

Recent research, including observations of large earthquakes,
geodetic deformation measurements, and numerical modeling have
raised the lower limits of the upper magnitude parameters in
subduction zones and confirmed that all major subduction zones
have essentially equal $b$-values and upper magnitude
parameters.
As magnitude records are broken in individual subduction
zones, the observed peak magnitudes are approaching the corner
magnitudes estimated from moment conservation.

Moment conservation implies that the corner magnitude of
subduction zones, taken as a group, should be well above
magnitude~9.
Given that subduction zones have statistically
indistinguishable parameters, magnitude~9 earthquakes can be
expected in any major subduction zone.
The global rate of magnitude~9 earthquakes, both predicted
from statistics of moderate events and from moment
conservation and observed during the last 110 years, is about
5 per century.

While earthquakes with a tapered form of Gutenberg-Richter
distribution and a corner magnitude of 9.6 would explain
observed tectonic deformation at plate boundaries, reasonable
models allow for even larger earthquakes.
Magnitude 10 earthquakes cannot be considered impossible, and
our models suggest a global recurrence time of a few hundred
or thousand years.

\section{Data and Resources }
\label{data}

The global CMT catalog of moment tensor inversions compiled by
the GCMT group is available at
http://www.globalcmt.org/CMTfiles.html
(last accessed December 2011).
The Centennial catalog by Engdahl and Villase\~nor (2002) is
available at
http://earthquake.usgs.gov/research/data/centennial.pdf
(last accessed December 2011).
Flinn-Engdahl Regions are explained and their coordinates as
well as {\scaps FORTRAN} files to process them are available
at
\vspace {-0.25 truecm}
\begin{small}\begin{verbatim}
http://earthquake.usgs.gov/learn/topics/flinn_engdahl_list.php
ftp://hazards.cr.usgs.gov/feregion/fe_1995/
\end{verbatim}\end{small}
\vspace {-0.25 truecm}
(last accessed December 2011).
{\sl Seismic Activity in Japan} Web site
\begin{small}\begin{verbatim}
http://www.jishin.go.jp/main/index-e.html
\end{verbatim}\end{small}
in the Knowledge section see items:
`Evaluation of Major Subduction-zone Earthquake;' and
`Probabilistic seismic Hazard map;'.
Also see
\begin{small}\begin{verbatim}
http://go.nature.com/yw5e92
\end{verbatim}\end{small}
(all sites last accessed May 2012).

\subsection* {Acknowledgments
}
\label{Ackn}

We are grateful to Peter Bird and Paul Davis of UCLA as well
as Robert Geller and Satoshi Ide of the Tokyo University for
useful discussion and suggestions.
We thank Kathleen Jackson who edited the manuscript.
Reviews by Seth Stein and Ross Stein as well as comments by
the Associate Editor Thorne Lay have been helpful in revising
the manuscript.
The authors appreciate partial support from the National
Science Foundation through grants EAR-0711515, EAR-0944218,
and EAR-1045876, as well as from the Southern California
Earthquake Center (SCEC).
SCEC is funded by NSF Cooperative Agreement EAR-0529922 and
USGS Cooperative Agreement 07HQAG0008.
Publication 0000, SCEC.

\pagebreak

\def\reference{\hangindent=22pt\hangafter=1}

\centerline { {\sc References} }
\vskip 0.1in
\parskip 1pt
\parindent=1mm
\def\reference{\hangindent=22pt\hangafter=1}

\reference
Anderson, J.\ G., 1979.
Estimating the seismicity from geological structure for seismic
risk studies,
{\sl Bull.\ Seismol.\ Soc.\ Amer.}, {\sl 69}, 135-158.

\reference
Annaka, T., Satake, K., Sakakiyama, T., Yanagisawa, K., and
Shuto, N. (2007),
Logic-tree approach for probabilistic tsunami hazard analysis
and its applications to the Japanese coasts,
{\sl Pure Appl. Geophys.}, {\bf 164}, 577-592.

\reference
Bateman, H., and Erdelyi, A., 1953.
{\sl Higher Transcendental Functions},
McGraw-Hill Co., NY.

\reference
Bird, P., 2010.
Can diligent and extensive mapping of faults provide reliable
estimates of the expected maximum earthquakes at these faults?
No.
AGU Fall Meet.\ Abstract S23B-02.

\reference
Bird, P., and Y. Y. Kagan, 2004.
Plate-tectonic analysis of shallow seismicity: apparent
boundary width, beta, corner magnitude, coupled
lithosphere thickness, and coupling in seven tectonic
settings,
{\sl Bull.\ Seismol.\ Soc.\ Amer.}, {\bf 94}(6), 2380-2399,
(plus electronic supplement),
see also an update at
\vspace {-0.25 truecm}
\begin{small}\begin{verbatim}
http://peterbird.name/publications/2004_global_coupling/2004_global_coupling.htm
\end{verbatim}\end{small}
\vspace {-0.25 truecm}

\reference
Brune, J.~N., 1968.
Seismic moment, seismicity, and rate of slip along major fault
zones,
{\sl J.\ Geophys.\ Res.}, {\bf 73}, 777-784.

\reference
Ekstr\"om, G., 2007.
Global seismicity: results from systematic waveform analyses,
1976-2005,
in {\sl Treatise on Geophysics}, {\bf 4}(4.16), ed.\ H.
Kanamori, pp.~473-481, Elsevier, Amsterdam.

\reference
Ekstr\"om, G., A. M. Dziewonski, N. N. Maternovskaya
and M. Nettles, 2005.
Global seismicity of 2003: Centroid-moment-tensor solutions
for 1087 earthquakes,
{\sl Phys.\ Earth planet.\ Inter.}, {\bf 148}(2-4), 327-351.

\reference
Engdahl, E.R. and Villase\~nor, A., 2002.
Global seismicity: 1900-1999,
in {\sl IASPEI Handbook of Earthquake and Engineering
Seismology}, W. H. K. Lee, H. Kanamori, P. C. Jennings, and C.
Kisslinger, Eds., {\bf part~A},
pp.~665-690, Boston, Academic Press, available at
\vspace {-0.25 truecm}
\begin{small}\begin{verbatim}
   http://earthquake.usgs.gov/research/data/centennial.pdf
\end{verbatim}\end{small}
\vspace {-0.25 truecm}

\reference
Fialko, Y., Sandwell, D., Simons, M. and Rosen, P., 2005.
Three-dimensional deformation caused by the Bam, Iran,
earthquake and the origin of shallow slip deficit,
{\sl Nature}, {\bf 435}, 295-299, doi:10.1038/nature03425.

\reference
Flinn, E.\ A., E.\ R.\ Engdahl, and A.\ R.\ Hill, 1974.
Seismic and geographical regionalization,
{\sl Bull.\ Seismol.\ Soc.\ Amer.}, {\bf 64}, 771-992.

\reference
Geller, R. J., 2011.
Shake-up time for Japanese seismology,
{\sl Nature}, {\bf 472}(7344), 407-409,
DOI: doi:10.1038/nature10105.

\reference
Grunewald, E. D. and R. S. Stein, 2006.
A new 1649-1884 catalog of destructive earthquakes near Tokyo
and implications for the long-term seismic process,
{\sl J. Geophys.\ Res.}, {\bf 111}, B12306,
doi:10.1029/2005JB004059.

\reference
Gutenberg, B., and C.\ F.\ Richter, 1954.
{\sl Seismicity of the Earth and Associated Phenomena},
Princeton, Princeton Univ.\ Press., 310~pp.

\reference
Hanks, T.C., 1992.
Small earthquakes, tectonic forces,
{\sl Science}, {\bf 256}, 1430-1432.

\reference
Hayes, G. P., P. S. Earle, H. M. Benz, D. J. Wald, and R. W.
Briggs the USGS/NEIC Earthquake Response Team, 2011.
88 Hours: The U.S. Geological Survey National Earthquake
Information Center Response to the 11 March 2011 Mw 9.0 Tohoku
Earthquake,
{\sl Seismol.\ Res.\ Lett.}, {\bf 82}(4), 481-493.

\reference
Headquarters for Earthquake Research Promotion, 2005.
National Seismic Hazard Maps for Japan available at
http://go.nature.com/yw5e92, also at
http://www.jishin.go.jp/main/index-e.html
(see Knowledge -- Probabilistic seismic Hazard map PDF 2,018
KB)

\reference
Jackson, D. D., and Y. Y. Kagan, 2011.
Characteristic earthquakes and seismic gaps,
In {\sl Encyclopedia of Solid Earth Geophysics},
Gupta, H. K. (Ed.), Springer, pp.~37-40,
DOI 10.1007/978-90-481-8702-7.

\reference
Kagan, Y.~Y., 1997.
Seismic moment-frequency relation for shallow earthquakes:
Regional comparison,
{\sl J. Geophys.\ Res.}, {\bf 102}(B2), 2835-2852.

\reference
Kagan, Y. Y., 2002a.
Seismic moment distribution revisited: I. Statistical results,
{\sl Geophys.\ J. Int.}, {\bf 148}(3), 520-541.

\reference
Kagan, Y. Y., 2002b.
Seismic moment distribution revisited: II. Moment conservation
principle, {\sl Geophys.\ J. Int.}, {\bf 149}(3), 731-754.

\reference
Kagan, Y. Y., 2002c.
Aftershock zone scaling,
{\sl Bull.\ Seismol.\ Soc.\ Amer.}, {\bf 92}(2), 641-655,
doi: 10.1785/0120010172.

\reference
Kagan, Y.~Y., 2005.
Earthquake slip distribution: A statistical model,
{\sl J. Geophys.\ Res.}, {\bf 110}(B5), B05S11,
doi:10.1029/2004JB003280, pp.~1-15 (with electronic
Appendices).

\reference
Kagan, Y. Y., P. Bird, and D. D. Jackson, 2010.
Earthquake patterns in diverse tectonic zones of
the globe,
{\sl Pure Appl.\ Geoph.}\ ({\sl The Frank Evison Volume}),
{\bf 167}(6/7), 721-741, doi: 10.1007/s00024-010-0075-3.

\reference
Kagan, Y.~Y., and D.~D.~Jackson, 1991.
Seismic gap hypothesis: ten years after,
{\sl J. Geophys.\ Res.}, {\bf 96}, 21,419-21,431.

\reference
Kagan, Y.~Y. and D.~D.~Jackson, 1999.
Worldwide doublets of large shallow earthquakes,
{\sl Bull.\ Seismol.\ Soc.\ Amer.}, {\bf 89}(5), 1147-1155.

\reference
Kagan, Y. Y., and D. D. Jackson, 2000.
Probabilistic forecasting of earthquakes,
{\sl Geophys.\ J. Int.}, {\bf 143}, 438-453.

\reference
Kagan, Y. Y. and Jackson, D. D., 2012.
Long- and short-term earthquake forecasts during the
Tohoku sequence,
manuscript, submitted to {\sl Earth, Planets and Space (EPS)},
EPS 2011/12/27, EPS3360TH2,
preprint http://arxiv.org/abs/1201.1659

\reference
Kaneko, Y. and Y. Fialko, 2011.
Shallow slip deficit due to large strike-slip earthquakes in
dynamic rupture simulations with elasto-plastic off-fault
response,
{\sl Geophys. J. Int.}, {\bf 186}, 1389-1403, doi:
10.1111/j.1365-246X.2011.05117.x

\reference
Koketsu, K., and Y. Yokota, 2011.
Supercycles along the Japan Trench and Foreseeability of the
2011 Tohoku Earthquake,
AGU Fall Meet.\ Abstract U33C-03.

\reference
Koravos, G.C., Tsapanos, T.M., and Bejaichund, M., 2006.
Probabilistic seismic hazard assessment for Japan,
{\sl Pure Appl.\ Geophys.}, {\bf 163}(1), 137-151, DOI:
10.1007/s00024-005-0003-0

\reference
Manighetti, I., Campillo, M., Sammis, C., Mai, P.M., and King,
G., 2005.
Evidence for self-similar, triangular slip distributions on
earthquakes: Implications for earthquake and fault mechanics,
{\sl J. Geophys.\ Res.}, {\bf 110}(B5), B05302.

\reference
Manighetti, I., Zigone, D., Campillo, M., and Cotton, F.,
2009.
Self-similarity of the largest-scale segmentation of the
faults: Implications for earthquake behavior,
{\sl Earth Planet.\ Sci.\ Lett.}, {\bf 288}(3-4), 370-381.

\reference
Matthews, M. V., W. L. Ellsworth, and P. A. Reasenberg, 2002.
A Brownian model for recurrent earthquakes,
{\sl Bull.\ Seismol.\ Soc.\ Amer.}, {\bf 92}, 2233-2250.

\reference
McCaffrey, R., 2007.
The Next Great Earthquake,
{\sl Science}, {\bf 315}, 1675,
DOI: 10.1126/science.1140173

\reference
McCaffrey, R., 2008.
Global frequency of magnitude 9 earthquakes,
{\sl Geology}, {\bf 36}(3), 263-266, DOI: 10.1130/G24402A.1
(GSA Data Repository item 2008063, Table DR1).

\reference
McGuire, J. J., and G. C. Beroza, 2012.
A Rogue Earthquake Off Sumatra,
{\sl Science}, {\bf 336}(6085), 1118-1119,
DOI: 10.1126/science.1223983.

\reference
Michael, A. J., 2011.
Random variability explains apparent global clustering of
large earthquakes,
{\sl Geophys. Res. Lett.}, {\bf 38}, L21301,
doi:10.1029/2011GL049443.

\reference
Minoura, K., F. Imamura, D. Sugawara, Y. Kono, and T.
Iwashita, 2001.
The 869 Jogan tsunami deposit and recurrence interval of
large-scale tsunami on the Pacific coast of northeast Japan,
{\sl J. Natural Disaster Sci.}, {\bf 23}, 83-88.

\reference
Molnar, P., 1979.
Earthquake recurrence intervals and plate tectonics,
{\sl Bull.\ Seismol.\ Soc.\ Amer.}, {\sl 69}, 115-133.

\reference
Nanjo, K. Z., H. Tsuruoka, N. Hirata, and T. H. Jordan, 2011.
Overview of the first earthquake forecast testing experiment
in Japan,
{\sl Earth Planets Space}, {\bf 63}(3), 159-169.

\reference
Nettles, M., Ekstr\"om, G., and H. C. Koss, 2011.
Centroid-moment-tensor analysis of the 2011 off the Pacific
coast of Tohoku Earthquake and its larger foreshocks and
aftershocks,
{\sl Earth Planets Space}, {\bf 63}(7), 519-523.

\reference
Nishenko, S. P., 1991.
Circum-Pacific seismic potential -- 1989-1999,
{\sl Pure Appl.\ Geophys.\ }, {\bf 135}, 169-259.

\reference
Ozawa, S., T. Nishimura, H. Suito, T. Kobayashi, M. Tobita, and
T. Imakiire, 2011.
Coseismic and postseismic slip of the 2011 magnitude-9
Tohoku-Oki earthquake,
{\sl Nature}, {\bf 475}, 373-376, DOI: doi:10.1038/nature10227

\reference
Rikitake, T. and I. Aida (1988).
Tsunami hazard probability in Japan,
{\sl Bull.\ Seismol.\ Soc.\ Amer.}, {\bf 78}, 1268-1278.

\reference
Ruff, L., and H. Kanamori, 1980.
Seismicity and the subduction process,
{\sl Phys.\ Earth Planet.\ Inter.}, {\bf 23}, 240-252.

\reference
Seismic Activity in Japan, 2008.
http://www.jishin.go.jp/main/index-e.html, see the Knowledge
section.

\reference
Shaw, B. E., and S. G. Wesnousky, 2008.
Slip-length scaling in large earthquakes: the role of
deep-penetrating slip below the seismogenic layer,
{\sl Bull.\ Seismol.\ Soc.\ Amer.}, {\bf 98}(4), 1633-1641.

\reference
Simons, M., Minson, S.E., Sladen, A., Ortega, F., Jiang, J.L.,
Owen, S.E., Meng, L.S., Ampuero, J.P., Wei, S.J., Chu, R.S.,
Helmberger, D.V., Kanamori, H., Hetland, E., Moore, A.W.,
and Webb, F.H., 2011.
The 2011 magnitude 9.0 Tohoku-Oki earthquake: mosaicking the
megathrust from seconds to centuries,
{\sl Science}, {\bf 332}(6036), 1421-1425 DOI:
10.1126/science.1206731.

\reference
Stein, S., Geller, R., and Liu, M., 2011.
Bad assumptions or bad luck: why earthquake
hazard maps need objective testing,
{\sl Seismol.\ Res.\ Lett.}, {\bf 82}(5), 623-626.

\reference
Stein, S., and E. A. Okal, 2007.
Ultralong Period Seismic Study of the December 2004 Indian
Ocean Earthquake and Implications for Regional Tectonics and
the Subduction Process,
{\sl Bull.\ Seismol.\ Soc.\ Amer.}, {\bf 97}(1A), S279-S295

\reference
Stein, S., and E. A. Okal, 2011.
The size of the 2011 Tohoku earthquake need not have been a
surprise,
{\sl Eos Trans.\ AGU}, {\bf 92}(27), 227-228.

\reference
Sugawara, D., F. Imamura, K. Goto, H. Matsumoto, and K.
Minoura, 2012.
The 2011 Tohoku-oki Earthquake Tsunami: Similarities and
Differences to the 869 Jogan Tsunami on the Sendai Plain,
{\sl Pure Appl. Geophys.}, DOI 10.1007/s00024-012-0460-1.

\reference
Uchida, N., and T. Matsuzawa, 2011.
Coupling coefficient, hierarchical structure, and earthquake
cycle for the source area of the 2011 off the Pacific coast of
Tohoku earthquake inferred from small repeating earthquake
data,
{\sl Earth Planets Space}, {\bf 63}(7), 675-679.

\reference
Wesnousky, S. G., C. H. Scholz, K. Shimazaki, and
T. Matsuda, 1984.
Integration of geological and seismological data for
the analysis of seismic hazard: a case study of Japan,
{\sl Bull.\ Seismol.\ Soc.\ Amer.}, {\bf 74}, 687-708.

\reference
Wyss, M., 1973.
Towards a physical understanding of the earthquake frequency
distribution,
{\sl Geophys.\ J.\ R.\ Astr.\ Soc.}, {\bf 31}, 341-359.

\reference
Young, J. B., B. W. Presgrave, H. Aichele, D. A. Wiens, and
E. A. Flinn, 1996.
The Flinn-Engdahl regionalisation scheme: the 1995 revision,
{\sl Phys.\ Earth Planet.\ Inter.}, {\bf 96}, 223-297.

\reference
Zaliapin, I. V., Y. Y. Kagan, and F. Schoenberg, 2005.
Approximating the distribution of Pareto sums,
{\sl Pure Appl.\ Geoph.}, {\bf 162}(6-7),
1187-1228.

\clearpage

\pagebreak

\newpage

\renewcommand{\baselinestretch}{1.25}

\setlength{\oddsidemargin}{-0.5cm}
\setlength{\evensidemargin}{-0.5cm}
\setlength{\textwidth}{17.0cm}


\begin{table}
\vspace {-1.5 truecm}
\caption{FE Subduction Seismic Zones, GCMT 1977-2010/12/31,
$m_t = 5.8$.}
\vspace{10pt}
\label{Table1}
\begin{tabular}{rrlrrrrrrr}
\hline
& & & & & & & & & \\[-15pt]
\multicolumn{1}{c}{}&
\multicolumn{1}{c}{FE}&
\multicolumn{1}{c}{Flinn-Engdahl }&
\multicolumn{1}{c}{}&
\multicolumn{1}{c}{}&
\multicolumn{1}{c}{$\dot M_T $}&
\multicolumn{1}{c}{}&
\multicolumn{1}{c}{}&
\multicolumn{1}{c}{$\dot M_s $}&
\multicolumn{1}{c}{}
\\[.3ex]
\multicolumn{1}{c}{No}&
\multicolumn{1}{c}{No}&
\multicolumn{1}{c}{seismic region name}&
\multicolumn{1}{r}{$N$}&
\multicolumn{1}{c}{$\beta \pm \sigma_\beta$}&
\multicolumn{1}{c}{$\times 10^{27}$}&
\multicolumn{1}{c}{$m_c \pm \sigma_M$}&
\multicolumn{1}{c}{$m_o$}&
\multicolumn{1}{c}{$\times 10^{27}$}&
\multicolumn{1}{c}{$\psi$}
\\[2pt]
\hline
  1 &  1 &Alaska-Aleutian Arc       &   280&  0.65$\pm$0.04&  5.10&  9.35$\pm$0.28&   8.0&  1.71& 0.336\\
  2 &  5 &Mexico-Guatemala          &   164&  0.60$\pm$0.06&  2.38&  9.17$\pm$0.29&   8.0&  1.35& 0.567\\
  3 &  6 &Central America           &   161&  0.68$\pm$0.06&  2.49&  9.22$\pm$0.29&   7.8&  0.59& 0.236\\
  4 &  7 &Caribbean Loop            &    59&  0.62$\pm$0.09&  1.05&  9.33$\pm$0.31&   7.4&  0.13& 0.125\\
  5 &  8 &Andean S. America         &   286&  0.57$\pm$0.04&  8.49&  9.74$\pm$0.28&   8.8&  9.54& 1.124\\
  6 & 12 &Kermadec-Tonga-Samoa      &   439&  0.80$\pm$0.04&  5.95&  9.12$\pm$0.28&   8.1&  2.16& 0.363\\
  7 & 13 &Fiji Is                   &    79&  0.85$\pm$0.10&  2.29&  9.72$\pm$0.30&   6.8&  0.06& 0.026\\
  8 & 14 &New Hebrides Is           &   424&  0.59$\pm$0.03&  4.81&  8.98$\pm$0.28&   7.9&  1.76& 0.366\\
  9 & 15 &Bismarck-Solomon Is       &   448&  0.60$\pm$0.03&  4.93&  8.95$\pm$0.28&   8.1&  2.46& 0.500\\
 10 & 16 &New Guinea                &   266&  0.66$\pm$0.05&  8.49&  9.80$\pm$0.28&   8.3&  1.48& 0.174\\
 11 & 18 &Guam-Japan                &    88&  0.86$\pm$0.10&  2.89&  9.82$\pm$0.30&   7.8&  0.34& 0.117\\
 12 & 19 &{\bf Japan-Kamchatka}     &   425&  {\bf 0.62$\pm$0.04}&  8.49&  {\bf 9.43$\pm$0.28}&   8.4&  5.30& 0.624\\
 12a & -- &{\bf Tohoku area} ($5^\circ \times 6^\circ$) & 109& {\bf 0.64$\pm$0.07}& 1.76& {\bf 9.26$\pm$0.29}& 7.7& 0.32& 0.180\\
 13 & 20 &S.E. Japan-Ryukyu Is      &    57&  0.62$\pm$0.10&  1.81&  9.79$\pm$0.31&   7.2&  0.10& 0.054\\
 14 & 21 &Taiwan                    &   110&  0.64$\pm$0.07&  1.53&  9.14$\pm$0.29&   7.7&  0.35& 0.226\\
 15 & 22 &Philippines               &   244&  0.68$\pm$0.05&  3.54&  9.17$\pm$0.28&   7.7&  0.74& 0.208\\
 16 & 23 &Borneo-Celebes            &   266&  0.68$\pm$0.05&  4.16&  9.23$\pm$0.28&   7.9&  1.09& 0.261\\
 17 & 24 &Sunda Arc                 &   278&  0.65$\pm$0.04&  6.51&  9.55$\pm$0.28&   8.6&  4.66& 0.716\\
 18 & 46 &Andaman Is-Sumatra        &   143&  0.71$\pm$0.07&  2.66&  9.37$\pm$0.29&   9.1& 15.26& 5.734\\
&&&&&&&&& \\
 & &1977-2010/12/31 ZONES          &  4217&  0.65$\pm$0.01& 77.57&  9.36$\pm$0.27&   9.1& 49.07& 0.633\\
&&&&&&&&& \\
 & &1977-1995/6/30 ZONES          &  2127 &  0.63$\pm$0.02& 27.40&  8.60$\pm$0.27&   8.4& 20.67& 0.754\\
\hline
\end{tabular}

\vskip 0.50cm
{\sl Notes:}
FE -- Flinn-Engdahl seismic region; W --
seismogenic width, $\mu$ -- elastic shear modulus, $\chi$ --
seismic coupling coefficient,
$N $ -- earthquake number,
$\beta $ -- parameter of the power-law distribution of
earthquake sizes,
$\dot M_T $ -- annual tectonic moment rate,
$\dot M_s $ -- annual seismic moment rate,
$m_c$ -- corner magnitude;
$m_o$ -- maximum moment magnitude observed in 1977-2010,
$\psi \, = \, \dot M_s/ \dot M_T $ -- ratio of seismic to
tectonic moment rate.
Seismic moment and moment rate are measured in {\sl dyne-cm}
and {\sl dyne-cm/yr}, respectively.
Tectonic rate for 1977-2010/12/31 period is calculated by
using Bird and Kagan (2004) parameters: $W$~=~104~km,
$\mu~=~49$~GPa, $\chi~=~0.5$.
In the last line of the Table we show the subduction zones
total calculation results for Kagan (1997, Table~1), where
the following parameters have been used: $W$~=~30~km,
$\mu~=~30$~GPa, $\chi~=~1.0$.
\hfil\break
\vspace{5pt}
\end{table}

\newpage

\renewcommand{\baselinestretch}{1.25}

\setlength{\oddsidemargin}{0.0cm}
\setlength{\evensidemargin}{0.0cm}
\setlength{\textwidth}{16.5cm}

\clearpage

\newpage

\renewcommand{\baselinestretch}{1.75}

\parindent=0mm

\begin{figure}
\begin{center}
\includegraphics[width=0.65\textwidth]{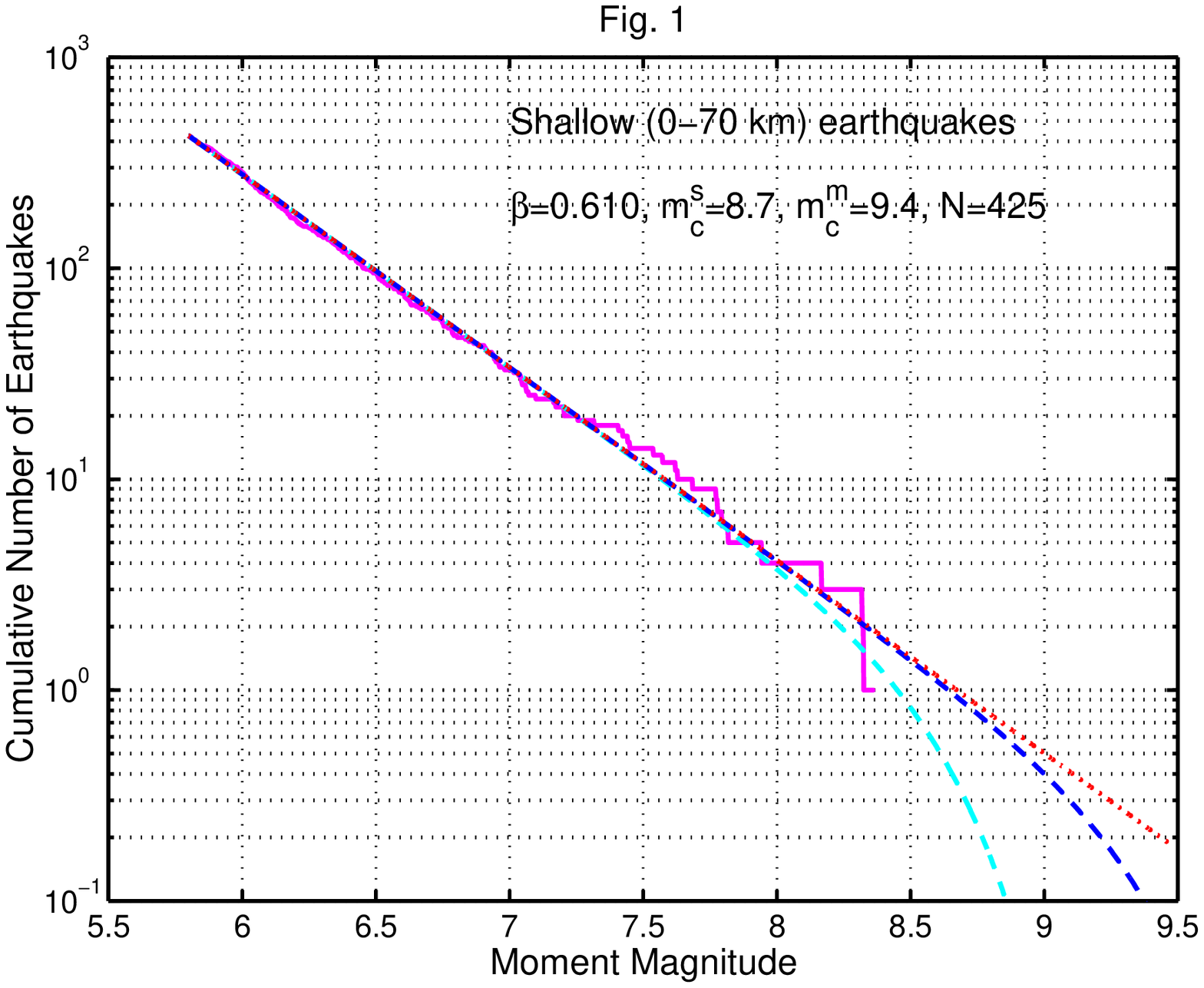}
%
\caption{\label{fig01}
}
\end{center}
Solid line -- the number of earthquakes in the Flinn-Engdahl
zone \#19 (Japan--Kurile-Kamchatka) with the moment magnitude
$(m)$ larger than or equal to $m$ as a function of $m$ for the
shallow earthquakes in the GCMT catalog during 1977--2010.
Magnitude threshold $m_t=5.8$, the total number of events is
425.
The unrestricted Gutenberg-Richter law is shown by a dotted
line (Kagan, 2002a).
Dashed lines show two tapered G-R (TGR) distributions:
the G-R law restricted at large magnitudes by an
exponential taper with a corner magnitude.
Left-hand line is for the corner magnitude $m_c^s=8.7$
evaluated by the maximum likelihood method using the
earthquake statistical record (with no upper limit, see
Fig.~\ref{fig02}).
Right-hand line is for the corner moment estimate $m_c^m=9.4$
is based on the moment conservation (see Table~\ref{Table1}).
The slope of the linear part of the curves corresponds to
$\beta=0.610$.
\end{figure}

\begin{figure}
\begin{center}
\includegraphics[width=0.75\textwidth]{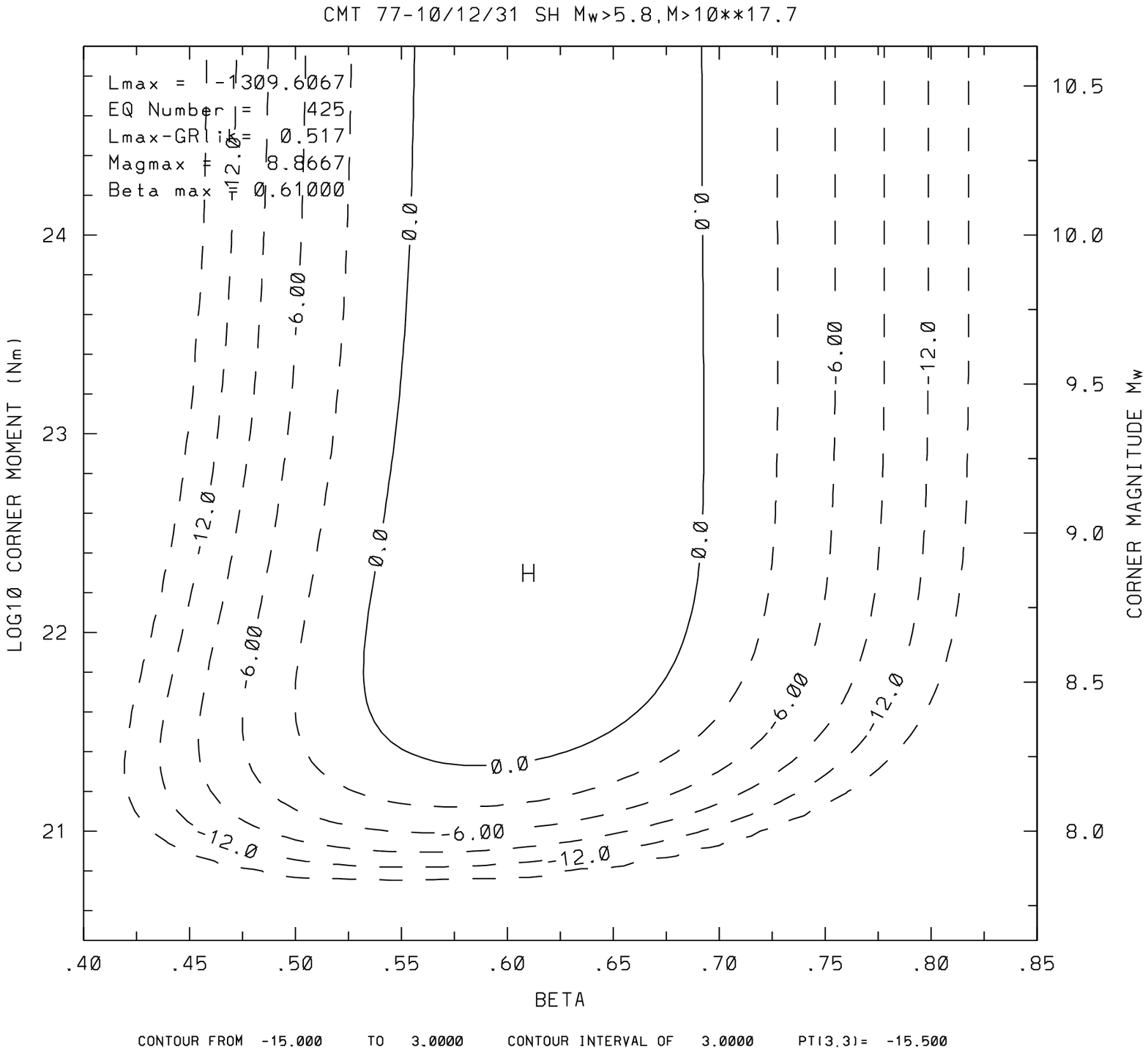}
%
\caption{\label{fig02}
}
\end{center}
A log-likelihood map for the distribution of the scalar
seismic moment of earthquakes in the Flinn-Engdahl zone \#19
(Japan--Kurile-Kamchatka):
the GCMT catalog time span is 1977/1/1--2010/12/31;
the seismic moment cutoff is $10^{17.7}$ Nm ($m_t = 5.8$);
the number of events is 425.
The approximation by the gamma distribution.
The `H' sign indicates the map maximum; the value of
the function is adjusted to be 3.0 at this place.
The zero contour (the solid line) corresponds to the 95\%
confidence area (Kagan, 1997).
\end{figure}

\begin{figure}
\begin{center}
\includegraphics[width=0.65\textwidth]{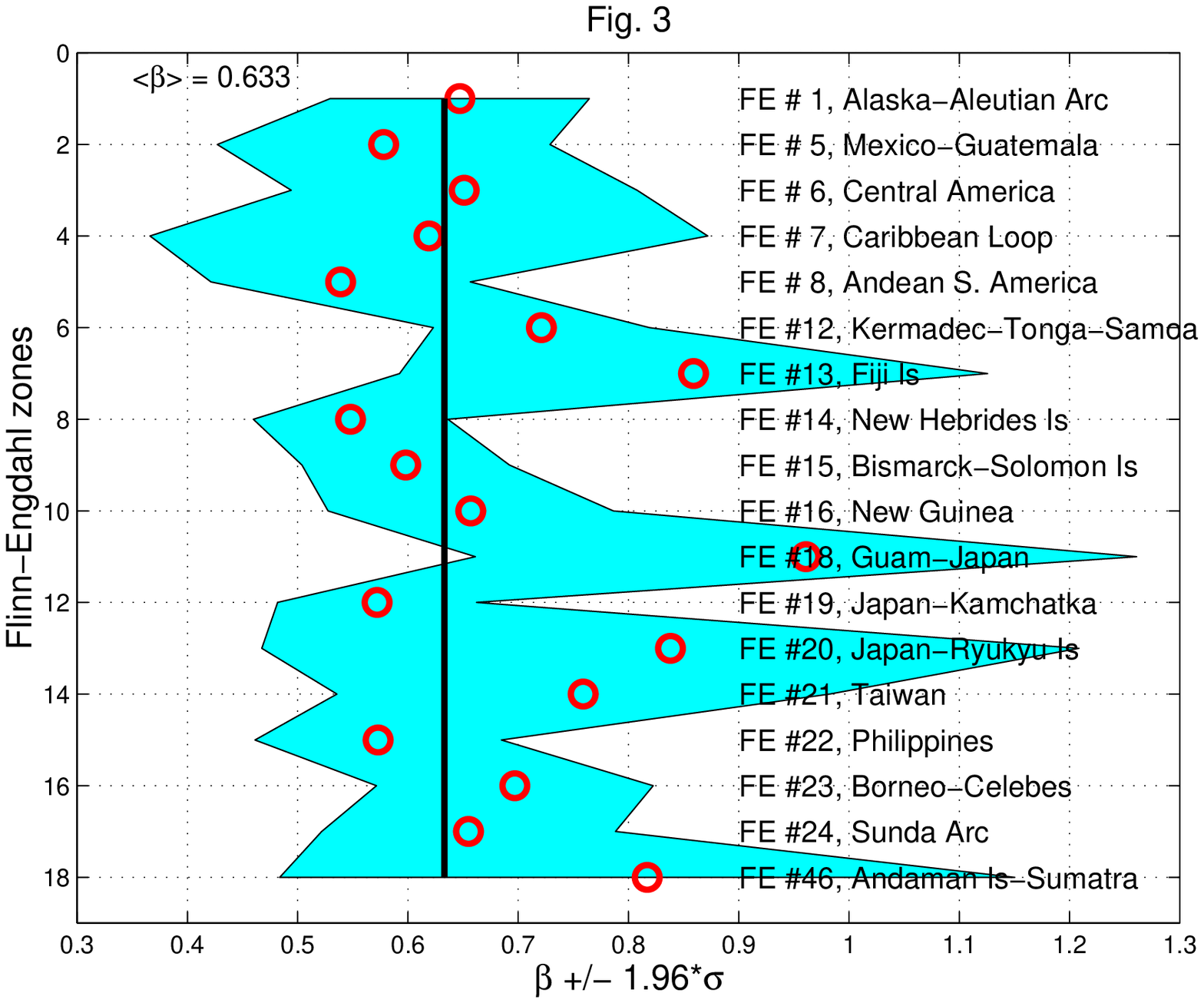}
%
\caption{\label{fig03}
}
\end{center}
Parameter $\beta$ distribution in the Flinn-Engdahl (FE)
subduction zones; average $\beta$-values are shown by circles.
GCMT catalog 1977-1995/6/30.
The ordinate numbers are sequential numbers of subduction
zones considered, the FE numbers and names for these zones are
shown in the right-hand part of the diagram.
Average region's $\beta$ and $\pm$1.96 standard deviations are
shown; the solid line corresponds to the average $<\beta>=
0.633$ for all subduction zones.
\end{figure}

\begin{figure}
\begin{center}
\includegraphics[width=0.65\textwidth]{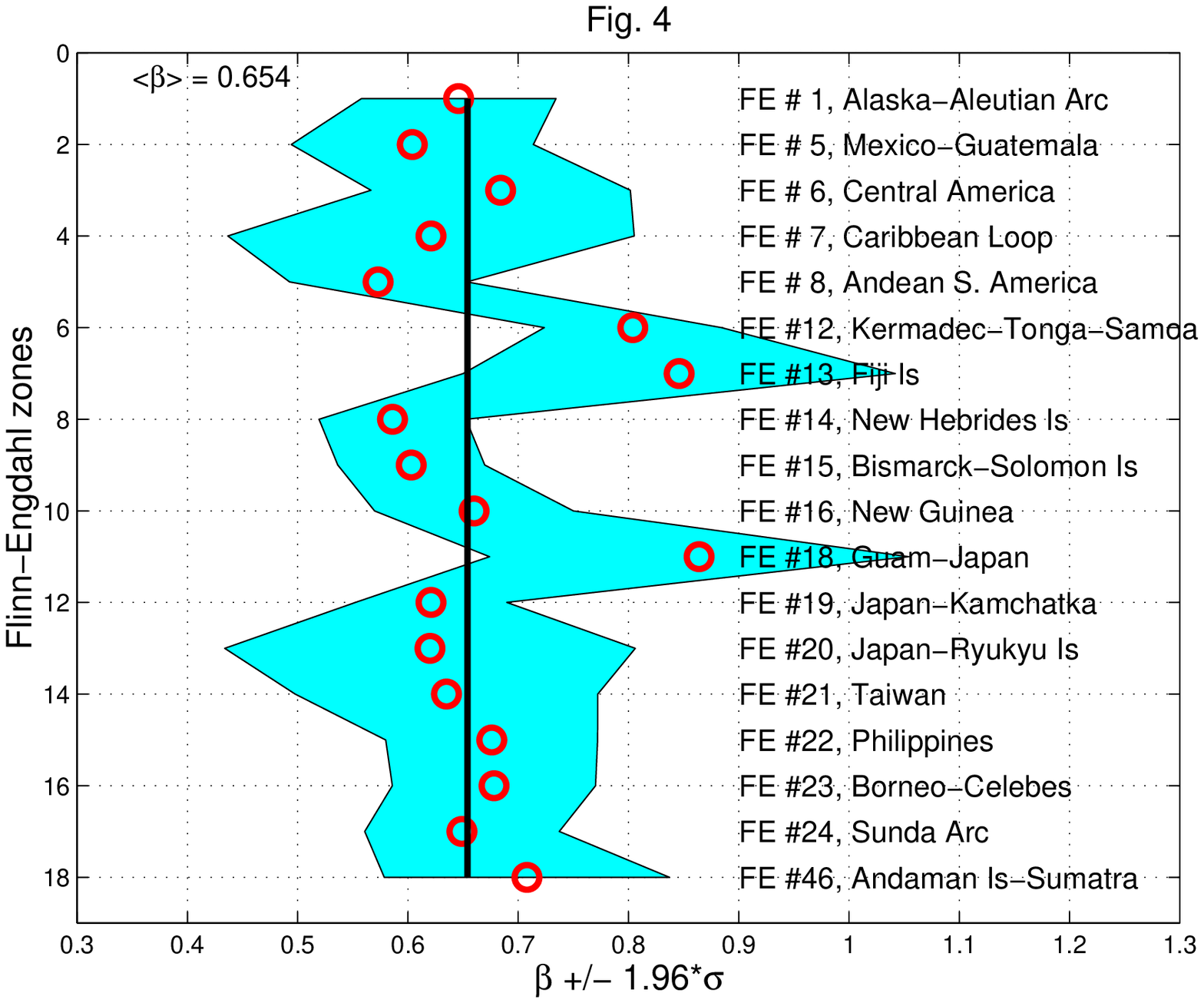}
%
\caption{\label{fig04}
}
\end{center}
Parameter $\beta$ distribution in Flinn-Engdahl subduction
zones.
GCMT catalog 1977-2010.
The average $<\beta>= 0.654$ for all subduction zones.
For notation see Fig.~\ref{fig03}.
\end{figure}

\begin{figure}
\begin{center}
\includegraphics[width=0.65\textwidth]{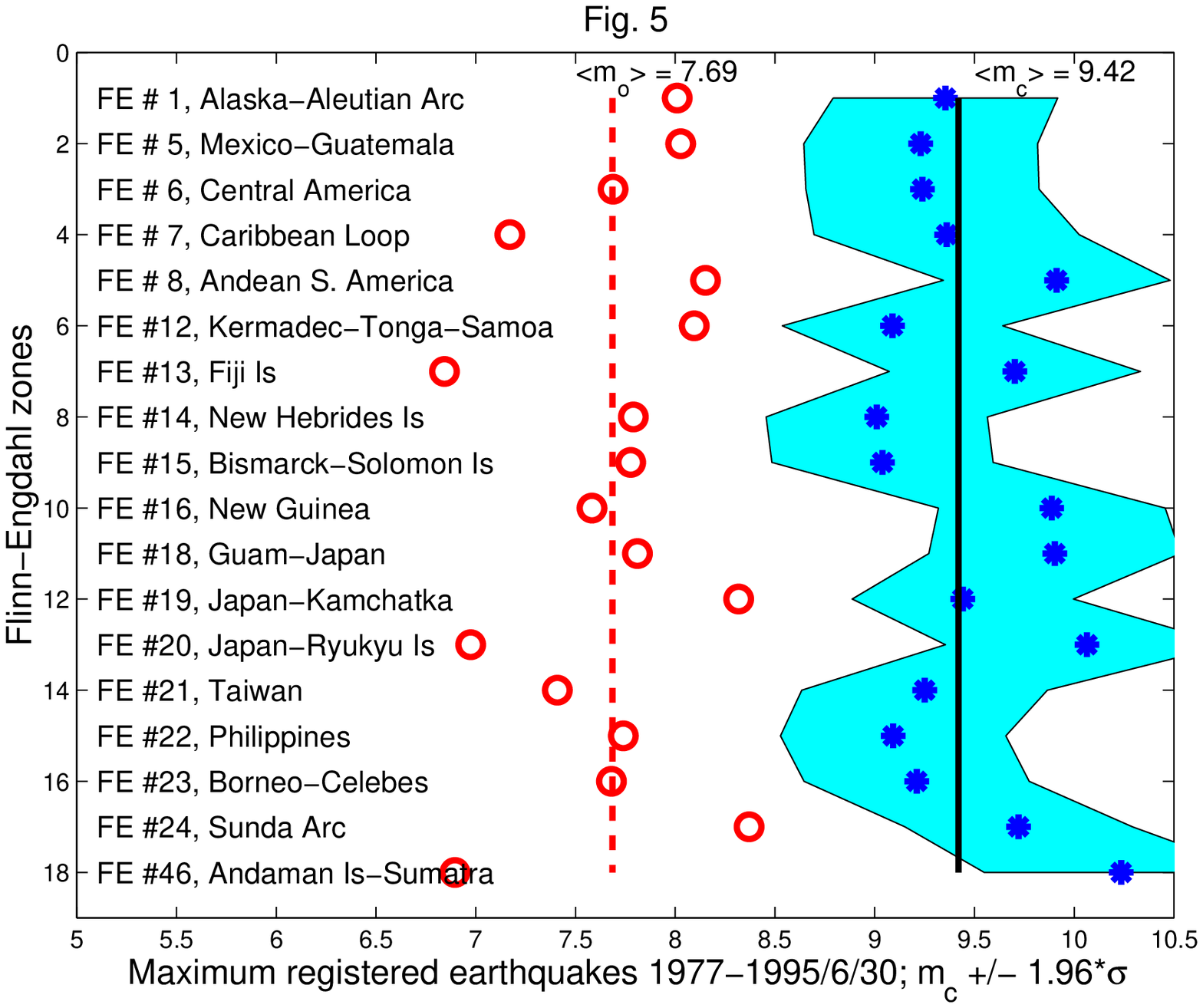}
%
\caption{\label{fig05}
}
\end{center}
The corner moment magnitude $m_c$ distribution in the
Flinn-Engdahl subduction zones.
GCMT catalog 1977-1995/6/30.
The region's $m_c$ and $\pm$1.96 standard deviations are
shown.
The solid line corresponds to the average $<m_c> = 9.42$ for
all subduction zones.
In $m_c$ calculations (Eq.~\ref{Eq06b}) we use the parameters
of the tectonic motion as proposed by Bird and Kagan (2004):
$W$~=~104~km, $\mu~=~49$~GPa, $\chi~=~0.5$.
Circles show events with the maximum magnitude $m_o$
in the regions during the catalog time interval.
The dashed line corresponds to the average $<m_o> = 7.69$ for
all subduction zones.
\end{figure}

\begin{figure}
\begin{center}
\includegraphics[width=0.65\textwidth]{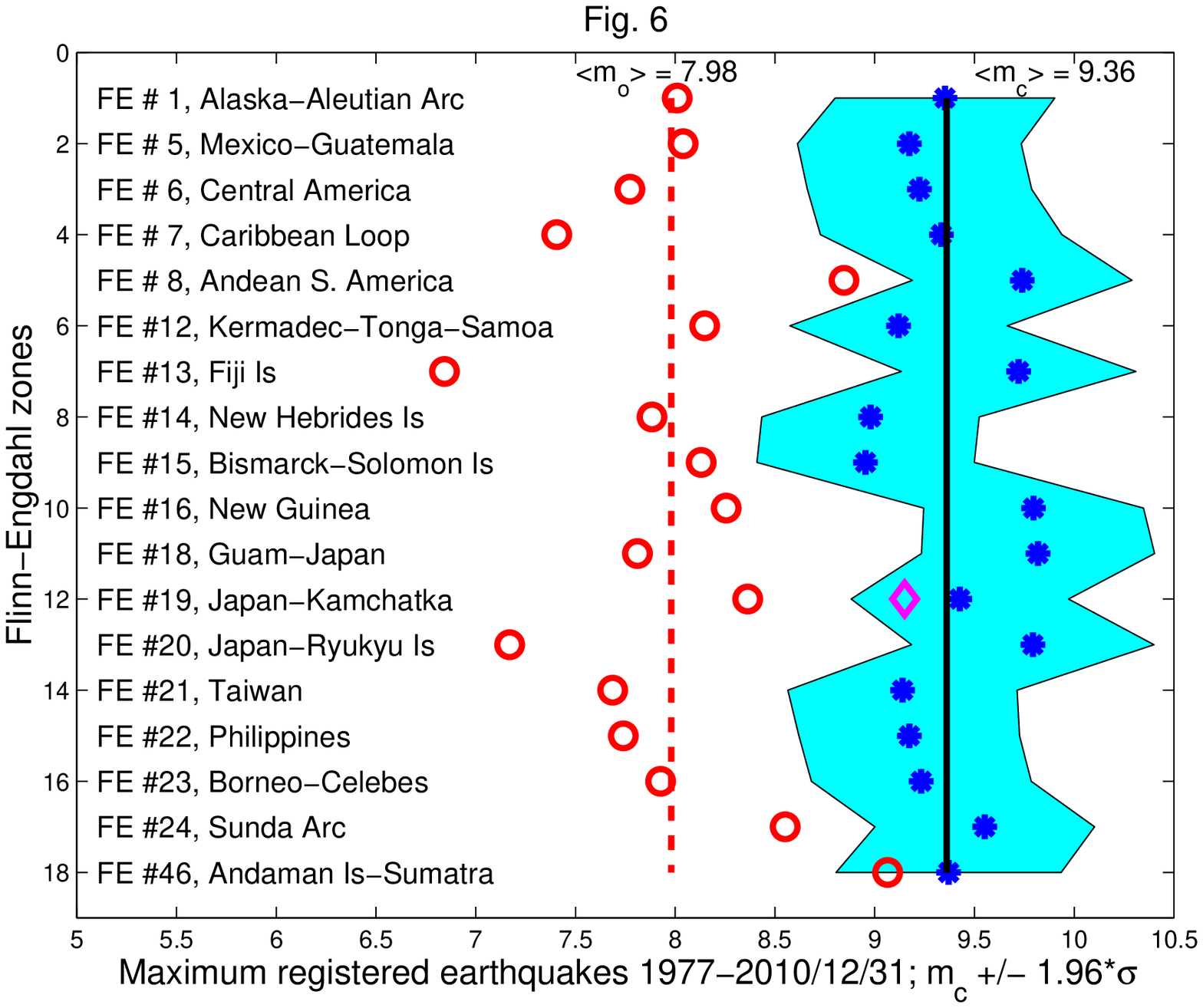}
%
\caption{\label{fig06}
}
\end{center}
The corner moment magnitude $m_c$ distribution in the
Flinn-Engdahl subduction zones.
GCMT catalog 1977-2010.
Averages of the corner magnitude ($<m_c> = 9.36$) and the
maximum observed magnitude ($<m_o> = 7.98$) for all subduction
zones are shown.
Diamond shows the Tohoku mega-earthquake magnitude.
For notation see Fig.~\ref{fig05}.
\end{figure}

\begin{figure}
\begin{center}
\includegraphics[width=0.65\textwidth]{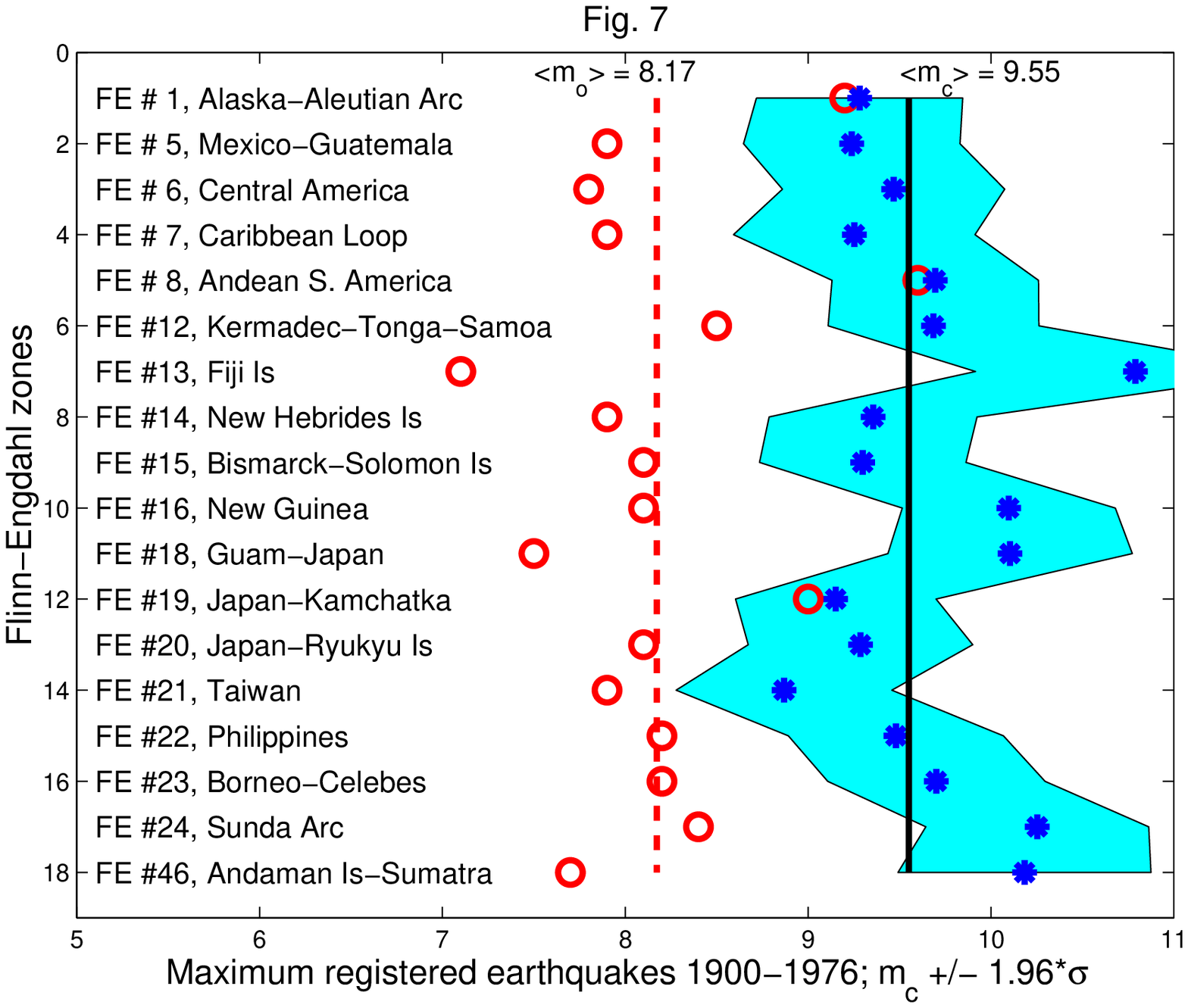}
%
\caption{\label{fig07}
}
\end{center}
The corner moment magnitude $m_c$ distribution in the
Flinn-Engdahl subduction zones, using the
Centennial catalog, 1900-1976, with magnitude threshold
$m_t=6.5$.
Averages $<m_c> = 9.55$ and $<m_o> = 8.17$ for all subduction
zones.
For notation see Fig.~\ref{fig05}.
\end{figure}

\begin{figure}
\begin{center}
\includegraphics[width=0.65\textwidth]{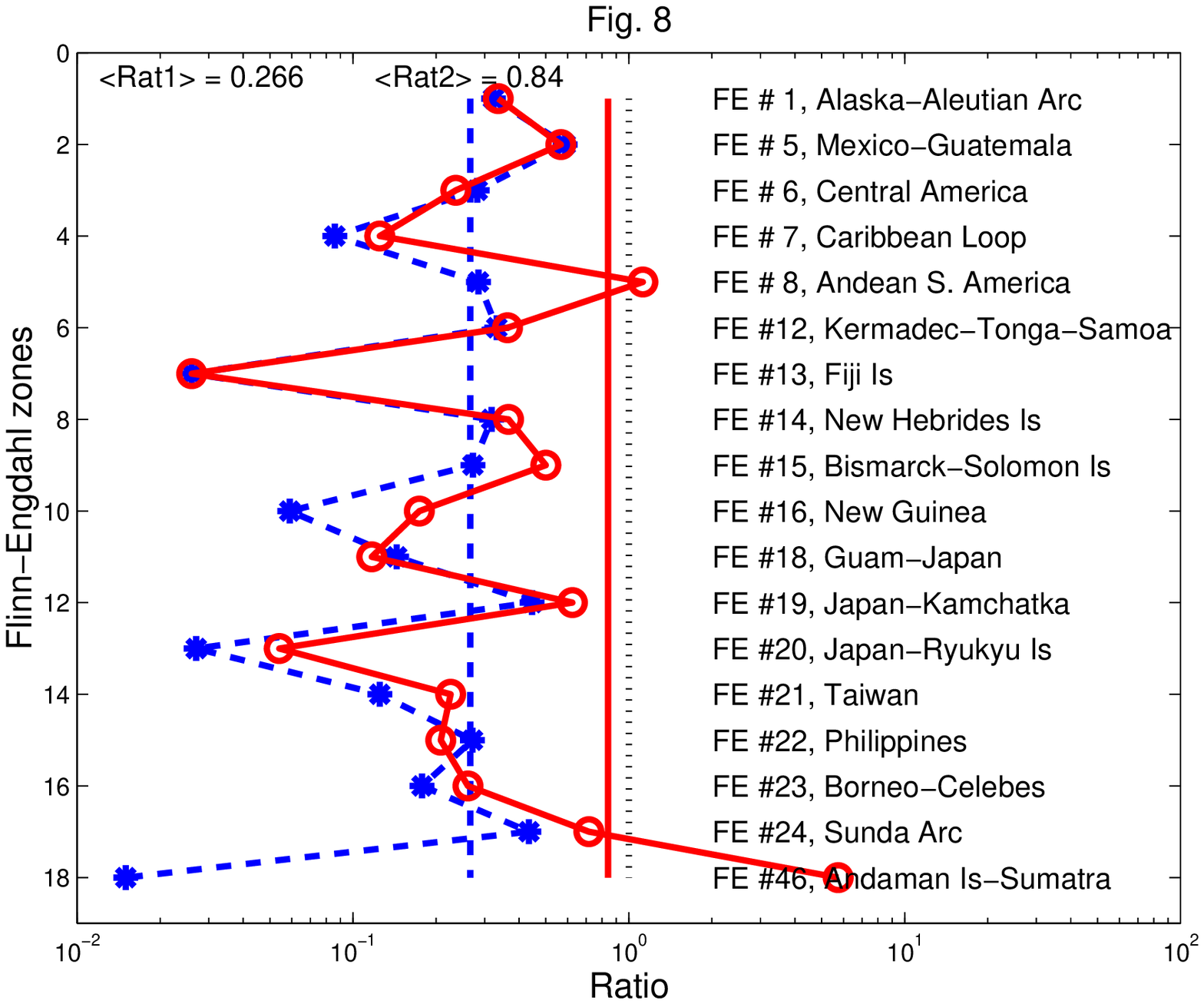}
%
\caption{\label{fig08}
}
\end{center}
Ratio ($\psi$) of the seismic to tectonic rate in the
Flinn-Engdahl subduction zones for the GCMT catalog
1977-1995/6/30 (dashed lines) and 1977-2010 (solid lines).
Vertical lines show the average ratio for all regions ($\psi_1
= 0.266$; $\psi_2 = 0.84$); the dotted line corresponds to the
ratio of 1.0.
See also the last column of Table~\ref{Table1}.
\end{figure}

\begin{figure}
\begin{center}
\includegraphics[width=0.65\textwidth]{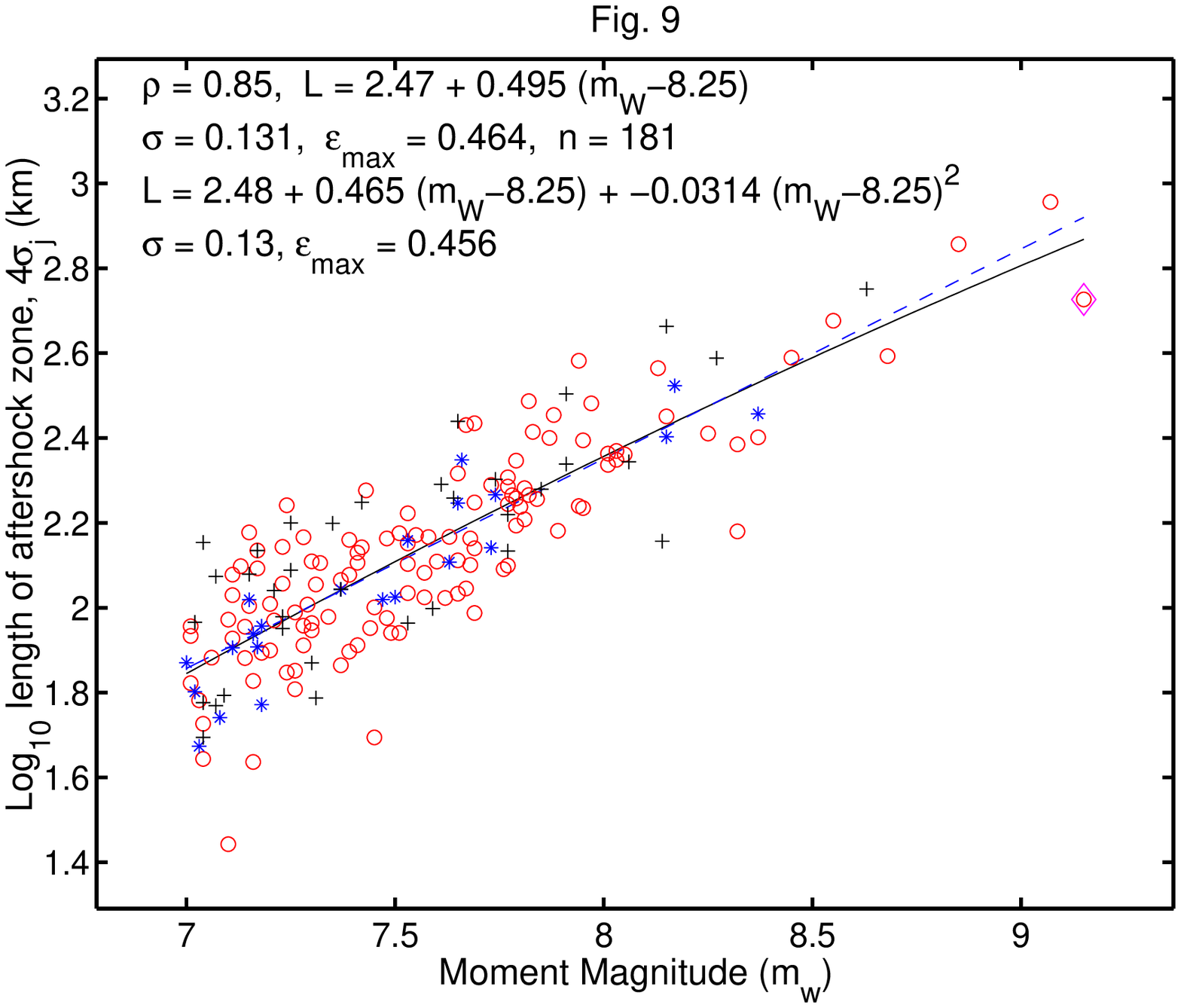}
%
\caption{\label{fig09}
}
\end{center}
The plot of the log aftershock zone length ($L$) against the
moment magnitude ($m$).
Earthquakes 1977-2012/04 are used.
The rupture length is determined using a 1-day aftershock
pattern.
The values of the correlation coefficient ($\rho$),
coefficients for linear (dashed line) and quadratic (solid
line) regression, standard ($\sigma$) and maximum
($\epsilon_{\rm max}$) errors, and the total number ($n$) of
aftershock sequences are shown in the diagram.
Both lines for the linear regression and the quadratic
approximation practically overlap in the plot.
\hfil\break
Circles -- thrust mainshocks;
\hfil\break
Stars -- normal mainshocks;
\hfil\break
Pluses -- strike-slip mainshocks.
\vskip .1in
The Tohoku mega-earthquake is marked by a
diamond sign.
\end{figure}

\newpage

\end{document}